\documentclass[12pt,a4paper]{article}
\usepackage{amsmath}
\usepackage{amssymb}
\usepackage{graphicx,epsfig}%ulem
\usepackage[mathscr]{eucal}
\usepackage{color} 
\usepackage[numbers,sort&compress]{natbib}
\usepackage[english]{babel}
\usepackage{tabularx}

\usepackage{setspace}
\usepackage{xspace}

\usepackage{multirow}

\newcommand\as{\alpha_{\mathrm{S}}} 
\newcommand\f[2]{\frac{#1}{#2}} 
 
\def\to{\rightarrow} 
\def\nn{\nonumber}

\def\msbar{{\overline {\rm MS}}} 

\def\T{{\bf T}}
\def\g{{\bf \Gamma}_t}
\def\vjk{v_{jk}}
\def\ep{\epsilon}
\def\F{{\bf F}_t^{(1)}}
\def\dcor{{\bf D}}

\def\rcut{r_{\rm cut}}

\def\ttbH{\ensuremath{t {\bar t}H}\xspace}
\def\ttb{\ensuremath{t {\bar t}}\xspace}

\usepackage{scalefnt,pstricks}
\usepackage{cancel}

\newcommand\Matrix{{\sc Matrix}\xspace}

\newcommand\OpenLoops{{\sc OpenLoops}\xspace}

\newcommand\madgraph{{\sc Madgraph5\_aMC@NLO}\xspace}

\usepackage{array}
\newcolumntype{L}[1]{>{\raggedright\let\newline\\\arraybackslash\hspace{0pt}}m{#1}}
\newcolumntype{C}[1]{>{\centering\let\newline\\\arraybackslash\hspace{0pt}}m{#1}}
\newcolumntype{R}[1]{>{\raggedleft\let\newline\\\arraybackslash\hspace{0pt}}m{#1}}

\usepackage{lipsum}
\usepackage[colorlinks=true,allcolors={blue!70!black}]{hyperref}

\begin{document} 
\begin{titlepage}
\begin{flushright}
ZU-TH 3/21\\
ICAS 061/21
\end{flushright}

\renewcommand{\thefootnote}{\fnsymbol{footnote}}
\vspace*{0.5cm}

\begin{center}
  {\Large \bf $\boldsymbol{\ttbH}$ production at NNLO:\\[0.2cm] the flavour off-diagonal channels}
\end{center}

\par \vspace{2mm}
\begin{center}
  {\bf Stefano Catani${}^{(a)}$, Ignacio Fabre${}^{(b,c)}$, Massimiliano Grazzini${}^{(b)}$,\\[0.2cm]
    }
and    
{\bf Stefan Kallweit${}^{(d)}$}

\vspace{5mm}

${}^{(a)}$INFN, Sezione di Firenze and
Dipartimento di Fisica e Astronomia,\\[0.1cm] 
Universit\`a di Firenze,
I-50019 Sesto Fiorentino, Florence, Italy\\[0.25cm]

${}^{(b)}$Physik Institut, Universit\"at Z\"urich, CH-8057 Z\"urich, Switzerland\\[0.25cm]

${}^{(c)}$ International Center for Advanced Studies (ICAS), ICIFI and ECyT-UNSAM, 25 de Mayo y Francia, (1650) Buenos Aires, Argentina

$^{(d)}$Dipartimento di Fisica, Universit\`{a} degli Studi di Milano-Bicocca and\\[0.1cm] INFN, Sezione di Milano-Bicocca,
I-20126, Milan, Italy\\[0.25cm]

\vspace{5mm}

\end{center}

\par \vspace{2mm}
\begin{center} {\large \bf Abstract} 

\end{center}
\begin{quote}
\pretolerance 10000

We consider QCD radiative corrections to the associated production of a heavy-quark pair ($Q{\bar Q}$) with a generic colourless system $F$ at hadron colliders.
We discuss the resummation formalism for the production of the $Q{\bar Q}F$ system at small values of its total transverse momentum $q_T$.
We present the results of the corresponding resummation coefficients at next-to-leading and, partly,
next-to-next-to-leading order.
The perturbative expansion of the resummation formula leads to the explicit ingredients that can be used to apply the $q_T$ subtraction formalism to fixed-order calculations for this class of processes.
We use the  $q_T$ subtraction formalism to perform a fully differential perturbative computation for the production of a top-antitop quark pair and a Higgs boson.
At next-to-leading order we compare our results
with those obtained with established subtraction methods and we
find complete agreement.
We present, for the first time, the results for the flavour off-diagonal partonic channels at the next-to-next-to-leading order.

\end{quote}

\vspace*{\fill}
\begin{flushleft}
February 2021
\end{flushleft}
\end{titlepage}

\section{Introduction}
\label{sec:intro}

The observation of Higgs boson production in association with a top quark--antiquark~(\ttb) pair was reported by the ATLAS and CMS collaborations in 2018~\cite{Aaboud:2018urx,Sirunyan:2018hoz}.
This production mode allows for a direct measurement of the top-quark Yukawa coupling.

The first theoretical studies of the production of a top--antitop quark pair and a Higgs boson~(\ttbH) were carried out in Refs.~\cite{Ng:1983jm,Kunszt:1984ri} at leading order~(LO) in QCD perturbation theory, and in Refs.~\cite{Beenakker:2001rj,Beenakker:2002nc,Reina:2001sf,Reina:2001bc,Dawson:2002tg,Dawson:2003zu} at next-to-leading order~(NLO).
NLO EW corrections were reported in Refs.~\cite{Frixione:2014qaa,Yu:2014cka,Frixione:2015zaa}.
The resummation of soft-gluon contributions close to the partonic kinematical threshold was considered in Refs.~\cite{Kulesza:2015vda,Broggio:2015lya,Broggio:2016lfj,Kulesza:2017ukk,Broggio:2019ewu,Kulesza:2020nfh}.
Full off-shell calculations with decaying top quarks were presented at NLO QCD~\cite{Denner:2015yca} and NLO QCD+EW~\cite{Denner:2016wet}.
The uncertainty of current theoretical predictions for \ttbH cross sections is at the ${\cal O}(10\%)$ level~\cite{deFlorian:2016spz}.
To match the experimental precision expected at the end of the high-luminosity phase of the LHC, next-to-next-to-leading order~(NNLO) predictions in QCD perturbation theory are required.

This paper is devoted to the NNLO (and NLO) QCD calculation of 
\ttbH production. 
At the partonic level, the NNLO calculation of \ttbH production requires 
the evaluation of tree-level contributions with two additional
unresolved partons in the final state, 
of one-loop contributions with one unresolved parton and 
of purely virtual contributions.
The required tree-level and one-loop scattering amplitudes can nowadays be evaluated with automated tools.
The two-loop amplitude for \ttbH production is not known. Being a five-leg amplitude involving particles with different masses, its computation is
at the frontier of current possibilities \cite{Heinrich:2020ybq}.

Even having all the required amplitudes, their implementation in a complete 
NNLO calculation at the fully differential (exclusive) level is
a highly non-trivial task because of the presence of infrared~(IR) 
divergences at intermediate stages of the calculation. 
In particular, these divergences do not permit a straightforward 
implementation of numerical techniques.
Various methods have been proposed and used to overcome these difficulties at the NNLO level (see Refs.~\cite{Bendavid:2018nar,Amoroso:2020lgh,TorresBobadilla:2020ekr,Heinrich:2020ybq} and references therein).

In this work we will use the transverse-momentum ($q_T$) subtraction method~\cite{Catani:2007vq}.
The $q_T$ subtraction formalism is
a method to handle and cancel the IR divergences in QCD computations at NLO, NNLO and beyond.
The method uses IR subtraction counterterms that are constructed by considering
and explicitly computing the $q_T$ distribution
of the produced final-state system in the limit $q_T \to 0$.
If the produced final-state system is composed of non-QCD (colourless) partons
(such as vector bosons, Higgs bosons, and so forth), the behaviour of the 
$q_T$ distribution in the limit $q_T \to 0$ has a universal 
(process-independent) structure that is explicitly known up to the NNLO level
through the formalism of transverse-momentum resummation~\cite{Catani:2013tia}.
These results on transverse-momentum resummation
are sufficient to fully specify the $q_T$ subtraction  formalism up to NNLO for this entire
class of processes\footnote{For this class of processes transverse-momentum resummation has recently been extended \cite{Li:2016ctv,Vladimirov:2016dll,Luo:2019szz,Ebert:2020yqt,Luo:2020epw} to next-to-next-to-next-to leading order (N$^3$LO), thus allowing
first N$^3$LO applications \cite{Cieri:2018oms,Billis:2019vxg} of the $q_T$ subtraction method.}.
The resummation formalism can, however, be extended to the production of final states containing a heavy-quark pair~\cite{Zhu:2012ts,Li:2013mia,Catani:2014qha}.
Exploiting such extension, the NNLO computations of top-quark and bottom-quark production were recently completed~\cite{Bonciani:2015sha,Catani:2019iny,Catani:2019hip,Catani:2020kkl}.

In this paper we consider the associated production of a top-quark pair with a Higgs boson.
Since the Higgs boson is colourless, the structure of transverse-momentum resummation for this process is closely analogous to that for heavy-quark production.
The only important difference is that
the emission of a Higgs boson off the top-quark pair changes the kinematics, and the top and antitop quarks are not back-to-back anymore at Born level.
We show that this feature can be controlled through the knowledge of appropriate resummation coefficients.
We present the explicit results of the resummation coefficients at NLO and, partly,
NNLO for the process of associated production of an
arbitrary number of heavy quark-antiquark pairs and a generic colourless system $F$ \cite{unpub}.
This allows us to obtain first results on the application of the 
$q_T$ subtraction method to the NLO and NNLO computations of \ttbH production
in hadron collisions.
We exploit the formulation of transverse-momentum resummation
in Ref.~\cite{Catani:2014qha} that includes the {\it complete} dependence on the
kinematics of the heavy-quark pair. This dependence and, in particular, the
complete control on the heavy-quark azimuthal correlations are essential (see Sec.~\ref{sec:formalism})
to extract all the NNLO counterterms of the $q_T$ subtraction method. 
Although the structure of transverse-momentum resummation for $\ttbH$
production is fully worked out up to NNLO, the explicit NNLO results
for the hard-virtual factors~\cite{Catani:2014qha} in the flavour diagonal
partonic channels $q{\bar q} \to t {\bar t}H+X$ and $gg \to t {\bar t}H+X$
($X$ denotes the unobserved inclusive final state)
are not yet known. Their evaluation requires the two-loop amplitudes for the partonic processes $q{\bar q} \to t {\bar t}H$ and $gg \to t {\bar t}H$,
as well as related soft contributions whose computation is available only for heavy-quark pair production.
Therefore, in the NNLO calculation of this paper
we present numerical results for all the flavour off-diagonal channels
$ab \to t {\bar t}H+X$, with $ab= qg ({\bar q}g), qq ({\bar q}{\bar q}),
qq' ({\bar q}{\bar q}'), q{\bar q}' ({\bar q} q')$ ($q$ and $q'$ denote quarks
with different flavours).

The paper is organised as follows. In Sect.~\ref{sec:formalism} we recall the transverse-momentum resummation formalism for the production of a high-mass system containing
a heavy-quark pair and discuss the perturbative ingredients needed for the NNLO calculation.
In Sect.~\ref{sec:resu} we present our numerical results for $ttH$ production at NLO and NNLO. In Sect.~\ref{sec:summa} we summarise our findings.
In the Appendix we report the explicit expressions of the resummation coefficients required for the calculation and highlight ensuing dynamical features related to azimuthal correlations and asymmetries.

\section[Transverse-momentum resummation and $q_T$ subtraction formalism for $Q{\bar QF}$ production]{Transverse-momentum resummation and $\boldsymbol{q_T}$ subtraction formalism for $\boldsymbol{Q{\bar Q}F}$ production}  
\label{sec:formalism}

We consider the
associated production of a heavy-quark pair ($Q{\bar Q}$) and an arbitrary colourless system $F$ in hadron collisions.
The system $F$ consists of one or more colourless particles, such as vector bosons, Higgs bosons, and so forth.
We denote by $M$ and ${\bf q_T}$ the invariant mass and transverse momentum of the $Q{\bar Q}F$ system, respectively.

At small values of $q_T$ (i.e., $q_T \ll M$) the perturbative QCD computation for this class of production processes is affected 
by large logarithmic terms of the type $\ln^n (M/q_T)$. These terms can formally be resummed to all perturbative orders.

In the case of $Q{\bar Q}$ production (i.e., no accompanying system $F$) the transverse-momentum resummation formalism
was developed in Ref.~\cite{Catani:2014qha} at arbitrary logarithmic accuracy, and by including the azimuthal-correlation contributions
(see also Refs.~\cite{Zhu:2012ts,Li:2013mia} for the azimuthally averaged case up to next-to-leading logarithmic accuracy).
The resummation formalism of Ref.~\cite{Catani:2014qha} can be extended to $Q{\bar Q}F$ associated production since the system $F$ 
is formed by colourless particles. The extension is straightforward at the formal level and, at the practical level, it requires the
explicit computation of process-dependent resummation factors at the necessary perturbative order.
In the following we briefly summarize the key steps and the ingredients that are involved in this extension.

Transverse-momentum resummation is performed through Fourier transformation from impact parameter space, where the impact
parameter vector ${\bf b}$ is the Fourier conjugated variable to the transverse-momentum vector ${\bf q_T}$.
The general resummation formula for both $Q{\bar Q}$ and $Q{\bar Q}F$ production is given in Eq.~(5) of Ref.~\cite{Catani:2014qha}.
The process-dependent contributions to this resummation formula are the LO partonic cross section
$\left[ d\sigma_{c{\bar c}}^{(0)} \right]$
($c=q,{\bar q},g$) and the resummation factor $\bigl( {\bf H \Delta } \bigr)$,
while all the other contributions are process independent.
These process-independent contributions and the corresponding resummation coefficients
are the same terms that control the production of a colourless high-mass system~\cite{Catani:2013tia} (see below).
The factor $\bigl( {\bf H \Delta } \bigr)$ has an all-order process-independent structure 
(see Eqs.~(10)--(16) and (26) in Ref.~\cite{Catani:2014qha}) that is controlled by 
resummation coefficients that can be explicitly computed at the required perturbative order.
These resummation coefficients are $(i)$ the soft anomalous dimension matrix ${\bf \Gamma}_t$,
$(ii)$ the radiative factor ${\bf D}$ and
$(iii)$ the subtraction operator 
$\widetilde{\bf I}_{c{\bar c}\to {\widetilde F}}$.

$(i)$ The resummation factor $\bf \Delta$ (see Eqs.~(15)--(18) in Ref.~\cite{Catani:2014qha}) depends on 
the soft anomalous dimension matrix ${\bf \Gamma}_t$, whose perturbative expansion in the QCD coupling $\as$ reads
\begin{equation}
  \label{eq:softan}
  {\bf \Gamma}_t(\as,\{p_i\})=\frac{\as}{\pi}\, {\bf \Gamma}_t^{(1)}(\{p_i\})+\left(\frac{\as}{\pi}\right)^2 {\bf \Gamma}_t^{(2)}(\{p_i\})+{\cal O}(\as^3) \;.
\end{equation}

$(ii)$ The term $\bf \Delta$ also depends on 
the radiative factor ${\bf D}({\bf \hat b},\as,\{p_i\})$,
which embodies azimuthal correlations of soft origin and, therefore, it depends on the direction ${\bf \hat b}$ of the impact parameter vector ${\bf b}$.
Its perturbative expansion reads
\begin{equation}
\label{eq:Dall}
{\bf D}({\bf \hat b},\as,\{p_i\}))  = 1+\frac{\as}{\pi}\, {\bf D}^{(1)}({\bf \hat b},\{ p_i\})+{\cal O}(\as^2) \;,
\end{equation}
with the constraint
\begin{equation}
  \label{eq:av}
  \langle {\bf D}({\bf \hat b},\as,\{p_i\}))\rangle_{\rm av.}  = 1\, ,
\end{equation}
where $\langle ...\rangle_{\rm av.}$ denotes the azimuthal average over ${\bf \hat b}$ (i.e., the average over the azimuthal angle ${\phi({\bf b})}$ of the 
transverse vector ${\bf b})$.

$(iii)$ The subtraction operator 
$\widetilde{\bf I}_{c{\bar c}\to {\widetilde F}}$
has the following perturbative expansion,
\begin{align}
\widetilde{\bf I}_{c{\bar c}\to {\widetilde F}}(\as(M^2), \ep;\{p_i\}) =
\sum_{n=1}^{\infty}
\left(\frac{\as(\mu_R^2)}{2\pi} \right)^{\!\!n}
\;\widetilde{\bf I}^{(n)}_{c{\bar c}\to {\widetilde F}}(\ep,M^2/\mu_R^2;\{p_i\})\;\;,
\label{eq:itilall}
\end{align}
where
$\mu_R$ is the renormalisation scale of the QCD coupling $\as(\mu_R^2)$.
Here ${\widetilde F}$ generically denotes the observed final-state system (i.e.,  ${\widetilde F}= Q{\bar Q}$ for heavy-quark pair production, 
or ${\widetilde F}= Q{\bar Q}F$ for the associated production process) with total invariant mass $M$.
The operator $\widetilde{\bf I}_{c{\bar c}\to {\widetilde F}}$ embodies IR-divergent contributions that are regularized 
by the customary procedure of analytic continuation in $d=4-2\ep$ space-time dimensions.
This subtraction operator contributes to the resummation factor $\bf H$ (see Eqs.~(12) and (13) in Ref.~\cite{Catani:2014qha}, and
Eqs.~(\ref{eq:Hq}) and (\ref{eq:Hg}) in the following) through the definition of the (IR-finite) hard-virtual amplitude
${\widetilde {\cal M}}_{c{\bar c} \to {\widetilde F}}$
(see Eq.~(26) in Ref.~\cite{Catani:2014qha} and Eq.~(\ref{eq:mtil}) in the following) of the partonic production process
$c{\bar c} \to {\widetilde F}$. 

The general transverse-momentum resummation formula for ${\widetilde F}= Q{\bar Q}, Q{\bar Q}F$ production involves 
a sole additional ingredient that is process dependent,
namely the scattering amplitude ${\cal M}_{c{\bar c} \to {\widetilde F}}$
of the partonic production process $c{\bar c} \to {\widetilde F}$. 

The resummation quantities ${\bf \Gamma}_t$, ${\bf D}$ and $\widetilde{\bf I}_{c{\bar c}\to {\widetilde F}}$ have
a `minimal' process dependence, which has a soft origin: they depend on the momenta $p_i$ and colour charges ${\bf T}_i$ of the 
colour-charged partons of the process $c{\bar c} \to {\widetilde F}$ (namely, the colliding partons $c$ and $\bar c$ 
and the produced heavy quarks and antiquarks). Such dependence is simply denoted by the argument $\{ p_i \}$
in Eqs.~(\ref{eq:softan}), (\ref{eq:Dall}) and (\ref{eq:itilall}). We also recall that 
${\bf \Gamma}_t$, ${\bf D}$ and $\widetilde{\bf I}_{c{\bar c}\to {\widetilde F}}$ are actually colour-space operators
that act on the colour indices of the corresponding partons.
The explicit expressions of the perturbative terms
${\bf \Gamma}_t^{(1)}$, 
${\bf \Gamma}_t^{(2)}$,
${\bf D}^{(1)}$ and $\widetilde{\bf I}_{c{\bar c}\to {\widetilde F}}^{(1)}$
for $Q{\bar Q}$ production were presented in Ref.~\cite{Catani:2014qha}, 
and we report the corresponding expressions for $Q{\bar Q}F$ production in the Appendix
of this paper.

According to the $q_T$ subtraction method~\cite{Catani:2007vq}, the formulation of transverse-momentum resummation for $Q{\bar Q}F$ production allows us
to write the (N)NLO partonic cross section $d{\sigma}^{Q{\bar Q}F}_{(N)NLO}$ as
\begin{equation}
\label{eq:main}
d{\hat \sigma}^{Q{\bar Q}F}_{(N)NLO}={\cal H}^{Q{\bar Q}F}_{(N)NLO}\otimes d{\hat \sigma}^{Q{\bar Q}F}_{LO}
+\left[ d{\hat \sigma}^{Q{\bar Q}F+\rm{jet}}_{(N)LO}-
d{\hat \sigma}^{Q{\bar Q}F, \, CT}_{(N)NLO}\right],
\end{equation}
where $d{\sigma}^{Q{\bar Q}F+\rm{jet}}_{(N)LO}$ is the $Q{\bar Q}F$+jet cross 
section 
at (N)LO accuracy.
To apply Eq.~(\ref{eq:main}) at NLO,  the LO cross section
$d{\sigma}^{Q{\bar Q}F+\rm{jet}}_{LO}$
can be directly obtained by integrating the corresponding tree-level scattering
amplitudes. To apply Eq.~(\ref{eq:main}) at NNLO,
$d{\sigma}^{Q{\bar Q}F+\rm{jet}}_{NLO}$
can be evaluated by using any available NLO method
to handle and cancel the corresponding IR divergences, if the relevant tree-level and one-loop QCD amplitudes are available.
Therefore, $d{\sigma}^{Q{\bar Q}F+\rm{jet}}_{(N)LO}$ is IR finite, {\it provided}
$q_T \neq 0$.
The square bracket term of Eq.~(\ref{eq:main}) is IR finite in the limit
$q_T \to 0$, but its individual contributions,
$d{\sigma}^{Q{\bar Q}F+\rm{jet}}_{(N)LO}$ and
$d{\sigma}^{Q{\bar Q}F, \, CT}_{(N)NLO}$, are separately divergent.
The IR-subtraction counterterm $d{\sigma}^{Q{\bar Q F}, \,CT}_{(N)NLO}$
is obtained from the (N)NLO perturbative expansion 
(see, e.g., Ref.~\cite{Bozzi:2005wk})
of the resummation formula
of the logarithmically enhanced
contributions to the corresponding $q_T$ distribution~\cite{Zhu:2012ts,Li:2013mia,Catani:2014qha}.
The explicit form of $d{\sigma}^{Q{\bar Q}F, \,CT}_{(N)NLO}$ 
can be completely worked out up to NNLO accuracy. It depends on the resummation coefficients that control
transverse-momentum resummation for the production of a colourless final-state system and, additionally, on the first two coefficients ${\bf \Gamma}_t^{(1)}$ and ${\bf \Gamma}_t^{(2)}$
of the soft anomalous dimension matrix in Eq.~(\ref{eq:softan}).

The explicit expression of the coefficient ${\bf \Gamma}_t^{(1)}$ for $Q{\bar Q} F$ production is given in the Appendix. The expression of ${\bf \Gamma}_t^{(2)}$ can be determined (see the Appendix)
by exploiting the relation~\cite{Catani:2014qha} between ${\bf \Gamma}_t$ and the IR singularities of the virtual scattering amplitude ${\cal M}_{c{\bar c}\to Q{\bar Q}F}$~\cite{Catani:2000ef, Mitov:2009sv, Ferroglia:2009ep, Ferroglia:2009ii, Mitov:2010xw}.

The IR-finite function ${\cal H}^{Q{\bar Q}F}$ in Eq.~(\ref{eq:main}) corresponds to the coefficient of the $\delta^{(2)}({\bf q_T})$ contribution in the expansion of the resummation formula.
It reads~\cite{Catani:2014qha}
\begin{equation}
  \label{eq:mainH}
  {\cal H}^{Q{\bar Q}F}_{c{\bar c};a_1a_2}=\langle \left[({\bf H}\, {\bf D})C_1C_2\right]_{c{\bar c};a_1a_2}\rangle_{\rm av.}\,,
\end{equation}
where the perturbative functions $C_1$ and $C_2$ are process independent and describe the emission of collinear radiation off the incoming partons.
In Eq.~(\ref{eq:mainH}) we have explicitly denoted the parton indices $\{c{\bar c},a_1,a_2\}$ that are implicit in Eq.~(\ref{eq:main}).
The indices $c$ and ${\bar c}$ correspond to the incoming partons of the LO partonic cross section $d{\hat \sigma}^{Q{\bar Q}F}_{LO}$.
The indices $a_1$ and $a_2$ are those of the parton densities $f_{a_1}$ and $f_{a_2}$ of the colliding hadrons.
The partonic cross section in Eq.~(\ref{eq:main}) depends on the renormalisation scale $\mu_R$ of $\as$ and on the factorisation scale $\mu_F$ of the parton densities.
    In Eq.~(\ref{eq:mainH}) and in 
    the following (see Eqs.~(\ref{eq:mainHq}) and (\ref{eq:mainHg}))
    the explicit structure of the function ${\cal H}^{Q{\bar Q}F}$ is presented by setting $\mu_R=\mu_F=M$.
The exact dependence on $\mu_R$ and $\mu_F$ can straightforwardly be recovered by using renormalisation group invariance and evolution of the parton densities.

In the quark annihilation channel ($c=q,{\bar q}$) the functions $C_1$ and $C_2$ do not depend on ${\bf b}$, and the symbolic factor $\left[({\bf H}\, {\bf D})C_1C_2\right]_{c{\bar c};a_1a_2}$ takes the form
\begin{equation}
  \left[({\bf H}\, {\bf D})C_1C_2\right]_{c{\bar c};a_1a_2}=\left({\bf H}\, {\bf D}\right)_{c{\bar c}} C_{ca_1}C_{{\bar c}a_2}~~~~~~(c=q,{\bar q})
\end{equation}
with 
\begin{equation}
  \label{eq:Hq}
  \left({\bf H}{\bf D}\right)_{c{\bar c}}=\frac{\langle \widetilde{\cal M}_{c{\bar c}\to Q{\bar Q}F}|{\bf D}|\widetilde{\cal M}_{c{\bar c}\to Q{\bar Q}F}\rangle}{\as^p(M^2)\,|{\cal M}_{c{\bar c}\to Q{\bar Q}F}^{(0)}(\{p_i\})|^2}~~~~~~(c=q,{\bar q})\, .
\end{equation}
The factor $\as^p|{\cal M}^{(0)}_{c{\bar c}\to Q{\bar Q}F}|^2$ in the denominator is the LO contribution to the squared amplitude $|{\cal M}_{c{\bar c}\to Q{\bar Q}F}|^2$ for the process $c{\bar c}\to Q{\bar Q}F$
(note that the power $p$ depends on the process, see Eq.~(\ref{eq:ampli})).
The IR-finite {\it hard-virtual} amplitude $\widetilde{\cal M}_{c{\bar c}\to Q{\bar Q}F}$ in Eq.~(\ref{eq:Hq})
is defined in terms of the all-order renormalised virtual amplitude ${\cal M}_{c{\bar c}\to Q {\bar Q}F}$
through an appropriate subtraction of IR singularities (see Eq.~(\ref{eq:mtil})).
By using Eq.~(\ref{eq:av}) the contribution of the azimuthal factor ${\bf D}$ to Eq.~(\ref{eq:mainH}) becomes trivial, and we obtain
\begin{equation}
  \label{eq:mainHq}
  {\cal H}^{Q{\bar Q}F}_{c{\bar c};a_1a_2}=\frac{\langle \widetilde{\cal M}_{c{\bar c}\to Q{\bar Q}F}|\widetilde{\cal M}_{c{\bar c}\to Q{\bar Q}F}\rangle}{\as^p(M^2)\,|{\cal M}_{c{\bar c}\to Q{\bar Q}F}^{(0)}(\{p_i\})|^2}\, C_{ca_1}\, C_{{\bar c}a_2}~~~~~~(c=q,{\bar q})\, .
\end{equation}

In the gluon fusion channel ($c=g$) the collinear functions $C_1$ and $C_2$
can be decomposed as~\cite{Catani:2010pd}
\begin{equation}
C_{ga}^{\mu\nu}(z,p_1,p_2,{\bf \hat b};\as)=d^{\mu\nu}(p_1,p_2)\,C_{ga}(z;\as)+D^{\mu\nu}(p_1,p_2,  {\bf \hat b})\,G_{ga}(z;\as)\, ,
\end{equation}
where the tensors $d^{\mu\nu}$ and $D^{\mu\nu}$, which multiply the helicity-conserving and helicity-flip components $C_{ga}$ and $G_{ga}$, read ($b^\mu=(0,{\bf b},0)$ with $b_\mu b^\mu=-{\bf b}^2$)
\begin{equation}
  d^{\mu\nu}(p_1,p_2)=-g^{\mu\nu}+\frac{p_1^\mu p_2^{\nu}+p_1^\nu p_2^{\mu}}{p_1\cdot p_2}\;,\qquad D^{\mu\nu}(p_1,p_2,{\bf \hat b})=d^{\mu\nu}(p_1,p_2)-2\frac{b^\mu b^\nu}{{\bf b}^2}\, .
\end{equation}
Therefore, the function ${\cal H}^{Q{\bar Q}F}_{gg;a_1a_2}$ reads
\begin{equation}
  \label{eq:mainHg}
  {\cal H}^{Q{\bar Q}F}_{gg;a_1a_2}=\langle({\bf H}\, {\bf D})_{gg;\mu_1\nu_1,\mu_2\nu_2}\, C_{ga_1}^{\mu_1\nu_1}({\bf \hat b}....)\, C_{ga_2}^{\mu_2\nu_2}({\bf \hat b}....)\rangle_{\rm av.}\,,
\end{equation}
where
\begin{equation}
\label{eq:Hg}
\left( {\bf H} \,{\bf D} \right)_{gg;\mu_1 \,\nu_1, \mu_2 \,\nu_2 }
=\f{\langle \,\widetilde{\cal M}_{gg 
\to Q {\bar Q}F}^{\nu_1^\prime \nu_2^\prime} \,|
\, {\bf D} \, | \,\widetilde{\cal M}_{gg 
\to Q {\bar Q}F}^{\mu_1^\prime \mu_2^\prime} \, \rangle
\;d_{\mu_1^\prime \mu_1} \;d_{\nu_1^\prime \nu_1}
\;d_{\mu_2^\prime \mu_2} \;
 d_{\nu_2^\prime \nu_2}
}{\as^p(M^2)\,|{\cal M}_{gg\to Q {\bar Q}F}^{(0)}(\{p_i\})|^2}\;\; .
\end{equation}

The functions $C_{ca}(z;\as)$ ($c=q,{\bar q},g$) and $G_{ga}(z;\as)$ have perturbative expansions
\begin{equation}
C_{ca}(z;\as)=\delta_{ca}\delta(1-z)+\sum_{n=1}^\infty \left(\frac{\as}{\pi}\right)^n C^{(n)}_{ca}(z)\;,\qquad G_{ga}(z;\as)=\sum_{n=1}^\infty \left(\frac{\as}{\pi}\right)^n G^{(n)}_{ga}(z)\, .
\end{equation}
The helicity-conserving coefficients $C^{(n)}_{ca}(z)$ are known up to $n=2$~\cite{Catani:2011kr,Catani:2012qa,Gehrmann:2012ze,Gehrmann:2014yya}, and
they are the same that contribute to Higgs
boson~\cite{Catani:2007vq} and vector-boson~\cite{Catani:2009sm} production.
Recently, their computation has been extended to the third order ($n=3$)~\cite{Luo:2019szz,Ebert:2020yqt,Luo:2020epw}.
The helicity-flip coefficients $G^{(n)}_{ga}(z)$ are known up to $n=2$~\cite{Gutierrez-Reyes:2019rug,Luo:2019bmw}.

The {\it hard-virtual} amplitude $\widetilde{\cal M}_{c{\bar c}\to Q{\bar Q}F}$ in Eq.~(\ref{eq:mainHq}) and (\ref{eq:Hg})
is expressed in terms of the all-order renormalised virtual amplitude ${\cal M}_{c{\bar c}\to Q {\bar Q}F}$ as
\begin{equation}
\label{eq:mtil}
| \,\widetilde{\cal M}_{c{\bar c}\to Q {\bar Q}F}(\{p_i\})  \rangle
= \left[ 1 -
\widetilde{\bf I}_{c{\bar c}\to Q {\bar Q}F}(\as(M^2), \ep;\{p_i\})
\right]
\;| {\cal M}_{c{\bar c}\to Q {\bar Q}F}(\{p_i\})  \rangle \;\;,
\end{equation}
where $\widetilde{\bf I}_{c{\bar c}\to Q {\bar Q}F}$ is the subtraction operator whose perturbative expansion is given in Eq.~(\ref{eq:itilall}).
The general expression of the first order coefficient $\widetilde{\bf I}^{(1)}_{c{\bar c}\to Q {\bar Q}F}$ in Eq.~(\ref{eq:itilall}) is known (see Appendix),
while the result for the second-order coefficient
is available only in the case of heavy-quark production~\cite{Angeles-Martinez:2018mqh,Catani:2019iny,inprep}.

The quantity
${\cal M}_{c{\bar c}\to Q {\bar Q}F}(\{ p_i \})$
on the right-hand side of  Eq.~(\ref{eq:mtil})
is the renormalised on-shell scattering amplitude and has the perturbative expansion
\begin{equation}
{\cal M}_{c{\bar c}\to Q {\bar Q}F}(\{ p_i \}) = \as^{p/2}(\mu_R^2) \,\mu_R^{p\ep}
\left[
{\cal M}_{c{\bar c}\to Q {\bar Q}F}^{\,(0)}(\{ p_i \})
\!+ \sum_{n=1}^{\infty} \left(\frac{\as(\mu_R^2)}{2\pi}\right)^{\!\!n}
\!\!{\cal M}_{c{\bar c}\to Q {\bar Q}F}^{\,(n)}(\{ p_i \}; \mu_R)
\right] .
\label{eq:ampli}
\end{equation}
The perturbative expansion of $\widetilde{\cal M}_{c{\bar c}\to Q {\bar Q}F}$
is completely analogous to that in Eq.~(\ref{eq:ampli}), with
$\widetilde{\cal M}_{c{\bar c}\to Q {\bar Q}F}^{\,(0)} =                                                                                                                                                                          
{\cal M}_{c{\bar c}\to Q {\bar Q}F}^{\,(0)}$ and the replacement
${\cal M}_{c{\bar c}\to Q {\bar Q}F}^{\,(n)} \to           
\widetilde{\cal M}_{c{\bar c}\to Q {\bar Q}F}^{\,(n)}$ ($n\geq 1$). At NLO we have
\begin{equation}
  \widetilde{\cal M}_{c{\bar c}\to Q {\bar Q}F}^{\,(1)}={\cal M}_{c{\bar c}\to Q {\bar Q}F}^{\,(1)}-\widetilde{\bf I}^{(1)}_{c{\bar c}\to Q {\bar Q}F} {\cal M}_{c{\bar c}\to Q {\bar Q}F}^{\,(0)}\, .
\end{equation}

Having discussed the perturbative ingredients entering the function ${\cal H}^{Q{\bar Q}F}_{c{\bar c};a_1a_2}$, we can now examine its NLO and NNLO expansions.
By inspecting Eqs.~(\ref{eq:mainHq}) and (\ref{eq:mainHg}) we see that the NLO truncation of ${\cal H}^{Q{\bar Q}F}_{c{\bar c};a_1a_2}$
receives contributions only from the tree-level and one-loop hard-virtual amplitudes ${\cal M}_{c{\bar c}\to Q {\bar Q}F}^{\,(0)}$ and $\widetilde{\cal M}_{c{\bar c}\to Q {\bar Q}F}^{\,(1)}$, and from the first-order
helicity-conserving coefficients $C_{ca}^{(1)}(z)$. Indeed, at this perturbative order the azimuthally dependent terms in ${\bf D}^{(1)}$ and $C^{\mu\nu}_{ga}$ do not contribute to Eq.~(\ref{eq:mainHg}) because of the azimuthal average.
The hard-virtual one-loop amplitude can be computed with available one-loop generators
such as \OpenLoops~\cite{Cascioli:2011va, Buccioni:2017yxi,Buccioni:2019sur} or {\sc Recola}~\cite{Actis:2012qn,Actis:2016mpe,Denner:2017wsf}.
As a consequence,  ${\cal H}^{Q{\bar Q}F}_{NLO}$ is
  available for the processes of interest, and, in particular, for \ttbH production.

The second-order coefficients 
${\cal H}^{Q{\bar Q}F}_{NNLO}$ are not available in general. Indeed, they depend on the hard-virtual amplitude $\widetilde{\cal M}_{c{\bar c}\to Q {\bar Q}F}^{\,(2)}$, which in turn requires the knowledge of
the renormalised two-loop amplitude, and of the subtraction operator $\widetilde{\bf I}^{(2)}_{c{\bar c}\to Q {\bar Q}F}$.
The computation of the two-loop amplitude for $Q{\bar Q}F$ production is at the frontier of current techniques, and, moreover, $\widetilde{\bf I}^{(2)}_{c{\bar c}\to Q {\bar Q}F}$ is also not known yet.
However, the NNLO contribution to ${\cal H}^{Q{\bar Q}F}_{c{\bar c};a_1a_2}$ can be completely determined
for all the flavour off-diagonal partonic channels $(a_1,a_2)\neq (c,{\bar c})$.
The perturbative ingredients entering the calculation in the quark--antiquark annihilation channel (see Eq.~(\ref{eq:mainHq})) are the corresponding tree-level (${\cal M}_{c{\bar c}\to Q {\bar Q}F}^{\,(0)}$)
and one-loop ($\widetilde{\cal M}_{c{\bar c}\to Q {\bar Q}F}^{\,(1)}$) hard-virtual amplitudes and the first- and second-order
helicity-conserving coefficients $C_{ab}^{(1)}(z)$ and $C_{ab}^{(2)}(z)$.
In the gluon fusion channel, 
the azimuthally dependent first-order coefficients of soft (${\bf D}^{(1)}$) and collinear ($G_{ga}^{(1)}$) origin are also required.
Indeed at NNLO such coefficients produce \cite{Catani:2014qha} non-vanishing mixed collinear--collinear and soft--collinear contributions in the expansion of Eq.~(\ref{eq:mainHg}).
The general expression of the first-order coefficient ${\bf D}^{(1)}({\bf \hat b},\{ p_i\})$ is explicitly known (see Appendix).
The evaluation of the ensuing NNLO contributions can be performed by computing the corresponding spin and colour-correlated squared tree-level amplitudes for the process $gg\to Q{\bar Q}F$.

In summary, the current knowledge of transverse-momentum resummation for high-mass systems
containing a heavy-quark pair and a colour singlet system allows us to use Eq.~(\ref{eq:main}) to obtain the complete NLO corrections for
this class of processes plus the NNLO corrections in the flavour off-diagonal partonic channels.

\section[Results for $t{\bar t}H$ production]{Results for $\boldsymbol{t{\bar t}H}$ production}
\label{sec:resu}

Having discussed the content of Eq.~(\ref{eq:main}), 
we are in a position to apply it to \ttbH production and to 
obtain the complete NLO results 
plus the NNLO corrections in all the flavour off-diagonal partonic channels. 
Our NLO implementation of the calculation has the main purpose of illustrating
the applicability of the $q_T$ subtraction method  to \ttbH production
and, in particular, of cross-checking the $q_T$ subtraction methodology by
numerical comparisons with NLO calculations performed by using more
established NLO methods. Our NNLO results on \ttbH production 
represent a first step (due to the missing flavour diagonal partonic channels)
towards the complete NNLO calculation for this production process.

Our results are obtained with two independent computations, which show complete agreement.
In the first computation, up to NLO, we use the phase space generation routines 
from the {\sc MCFM} program~\cite{Campbell:2015qma}, 
suitably modified for $q_T$ subtraction along the
lines of the corresponding numerical programs for 
Higgs boson~\cite{Catani:2007vq} and vector-boson~\cite{Catani:2009sm} 
production. The numerical integration is carried out using the {\sc Cuba} library~\cite{Hahn:2004fe}.
At NNLO accuracy the \ttbH+jet cross section is evaluated by using 
the {\sc Munich} code\footnote{{\sc Munich}, which is the abbreviation of “MUlti-chaNnel Integrator at Swiss (CH) precision”, is an automated parton-level NLO generator by S. Kallweit.}, which
provides a fully automated implementation of the NLO dipole subtraction formalism 
\cite{Catani:1996jh,Catani:1996vz,Catani:2002hc} and an efficient phase space integration. The remaining flavour off-diagonal contributions at NNLO are evaluated with a dedicated fortran implementation.
The second computation is directly implemented within the \Matrix framework \cite{Grazzini:2017mhc}, suitably extended to \ttbH production.
In both implementations all the required tree-level and one-loop amplitudes are obtained with \OpenLoops~\cite{Cascioli:2011va, Buccioni:2017yxi,Buccioni:2019sur},
including the tree-level spin- and colour-correlated amplitudes required to evaluate the contributions in Eq.~(\ref{eq:mainHg}).

In order to  numerically evaluate the contribution in the square bracket of Eq.~(\ref{eq:main}), a technical cut-off $\rcut$ is introduced on the dimensionless variable $q_T / M$, where $M$ is the invariant mass of the $t\bar t H$ system. The final result, which corresponds to the limit $\rcut \to 0$, is extracted by computing the cross section at fixed values of $\rcut$ in the range $[0.01\%, r_{max}]$. Quadratic least $\chi^2$ fits are performed for different values of $r_{max}\in[0.5\%, 1\%]$. The extrapolated value is then extracted from the fit with lowest $\chi^2/$degrees-of-freedom, and the uncertainty is estimated by comparing the results obtained by the different fits. This procedure is the same as implemented in {\sc matrix}~\cite{Grazzini:2017mhc} and it has been shown to provide a conservative estimate of the systematic uncertainty in the $q_T$ subtraction procedure for various processes (see Sec. 7 in Ref.~\cite{Grazzini:2017mhc}). 

%%====================================
\renewcommand\arraystretch{1.3}
\begin{table}[t]
\centering
\begin{tabular}{l|r|r}
$\sigma$ [fb] & \multicolumn{1}{c|}{$13~\mathrm{TeV}$} & \multicolumn{1}{c}{$100~\mathrm{TeV}$} \\
\hline\hline
LO & $394.987(3)\phantom{000}$  & $28228.2(2)\phantom{.000}$\\
\hline
NLO ({\sc Madgraph5\_aMC@NLO}) & {$499.76(4)\phantom{0000}$} & $36948(3)\phantom{..0000}$\\
NLO (\Matrix) & $499.73(1)\phantom{0000}$ & $36947(1)\phantom{..0000}$\\
NLO ($q_T$) & $499.79(4)\phantom{0000}$ & $36947(3)\phantom{..0000}$\\
\hline
${\cal O}(\as^4)_{qg}$ & $-0.796(65)\phantom{00}$ & $218.3(5.0)\phantom{00}$ \\
${\cal O}(\as^4)_{q({\bar q})q^\prime}$ & $0.62694(82)$ & $95.307(56)\phantom{.}$\\
\end{tabular}
\vspace*{1ex}
\caption{The $ttH$ total cross section at LO and NLO, and its NNLO corrections
in the flavour off-diagonal partonic channels.
The numerical uncertainties at LO and NLO (\madgraph, {\sc Matrix})
are due to numerical integration, while at NLO ($q_T$ subtraction) and NNLO they
also include the systematics uncertainty from the $\rcut\to 0$ extrapolation.}
\label{tab:results}
\end{table}
%%====================================

We consider $pp$ collisions at the centre-of-mass energies $\sqrt{s}=13~\mathrm{TeV}$ and $\sqrt{s}=100~\mathrm{TeV}$.
We use the NNPDF31~\cite{Ball:2017nwa} parton distribution functions (PDFs) 
with the QCD running coupling $\as$ evaluated
at each corresponding order
(i.e., we use $(n+1)$-loop $\as$ at N$^n$LO, with $n=1,2$).
The pole mass of the top quark is $m_t=173.3$~GeV, the Higgs boson mass $m_H=125$ GeV, and the Fermi constant $G_F = 1.16639\times 10^{-5}$\,GeV$^{-2}$.
  The renormalisation and
factorization scales, $\mu_R$ and $\mu_F$, are fixed at
$\mu_R=\mu_F=(2m_t+m_H)/2$.
Our predictions for the LO and NLO cross sections and for the NNLO corrections in the flavour off-diagonal channels are presented in Table~\ref{tab:results}
together with their uncertainties due to the numerical integration and the extrapolation to $\rcut\to 0$, computed as explained above.
The NLO cross section computed with $q_T$ subtraction is compared with the result obtained with \madgraph{}~\cite{Alwall:2014hca}, which uses FKS subtraction \cite{Frixione:1995ms,Frixione:1997np} and with the corresponding result obtained with \Matrix,
which implements dipole subtraction \cite{Catani:1996jh,Catani:1996vz,Catani:2002hc}.

\begin{figure}[t]
\begin{center}
\includegraphics[width=0.49\textwidth]{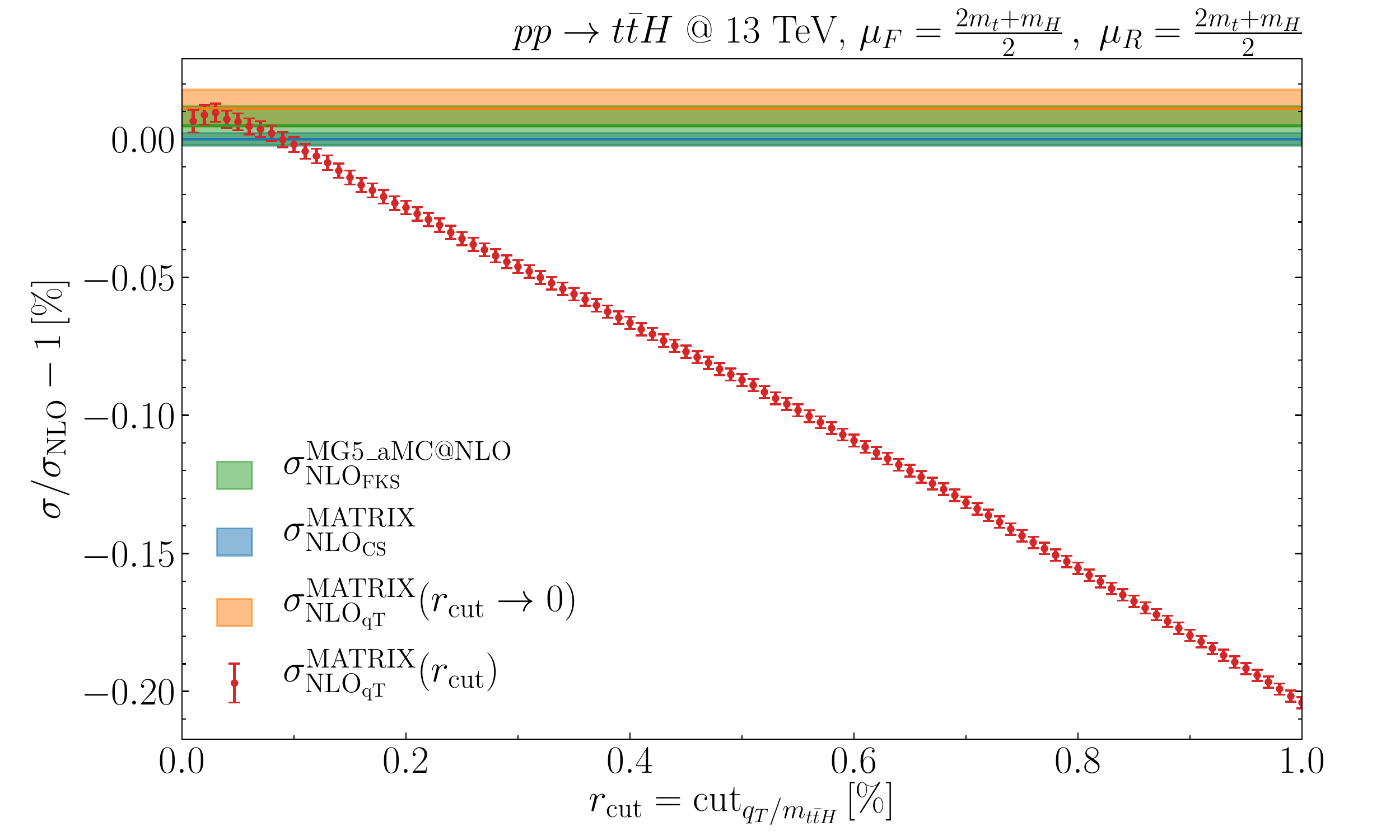}
\includegraphics[width=0.49\textwidth]{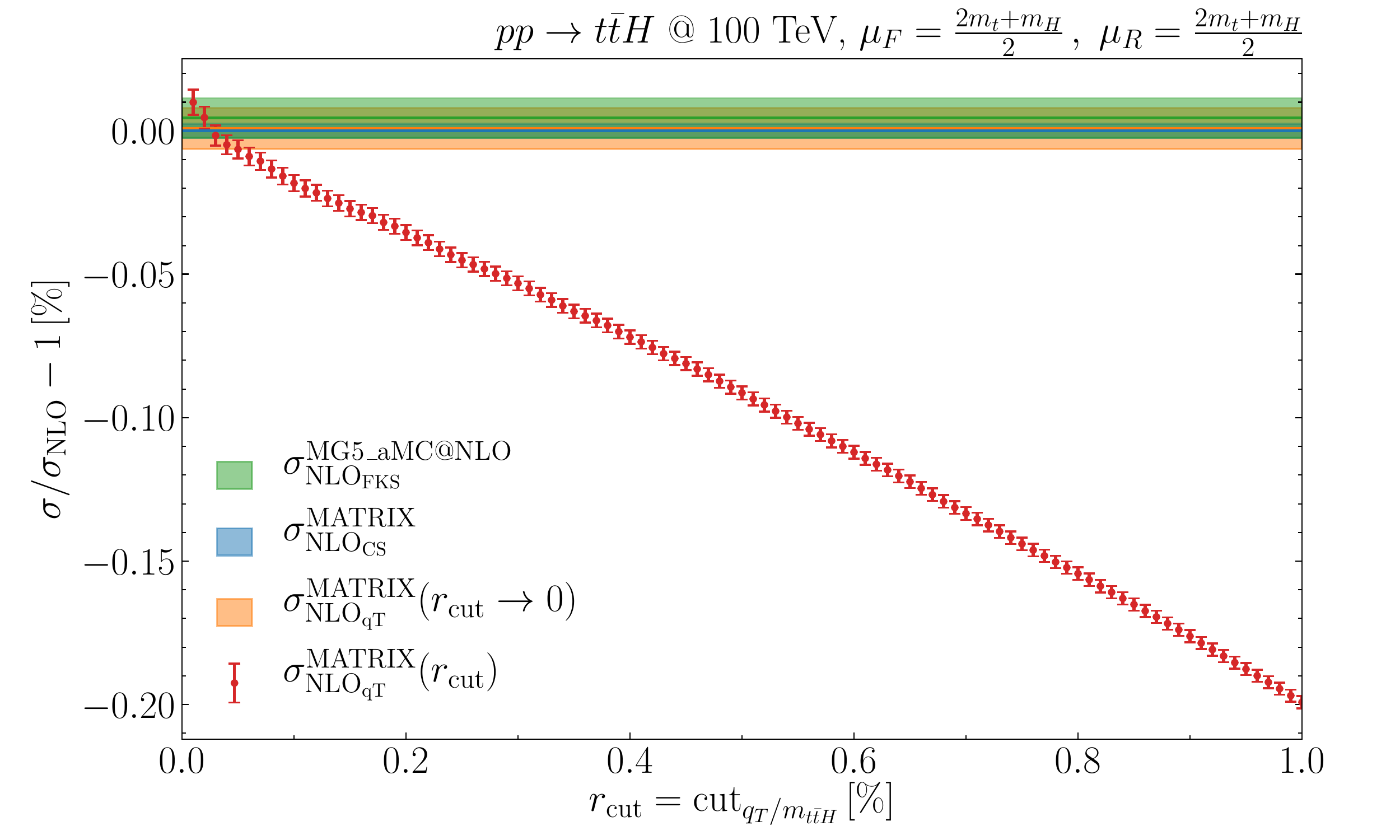}
\end{center}
\vspace{-1ex}
\caption{The $\rcut$ dependence (data points) at $\sqrt{s}=13~\mathrm{TeV}$ (left) and $100~\mathrm{TeV}$ (right) of the NLO total cross section computed by using $q_T$ subtraction.
The bands show the extrapolated value at $\rcut \to 0$ and the NLO results
from \madgraph (using FKS subtraction) and {\sc Matrix} (using dipole subtraction).}
\label{fig:r_cut}
\end{figure}

We start our discussion from the NLO results. The NLO corrections increase the LO result by $27\%$ ($31\%$) at $\sqrt{s}=13$ TeV ($\sqrt{s}=100$ TeV).
The flavour off-diagonal $qg+{\bar q}g$ channel contributes about $15\%$ ($23\%$) of the total NLO correction. 
As expected, from Table~\ref{tab:results} we observe excellent agreement between the NLO cross section obtained with \madgraph and \Matrix.
The result obtained with $q_T$ subtraction also agrees with \madgraph and \Matrix results.
The quality of the $\rcut\to 0$ extrapolation can be assessed by studying the behavior of the cross section at fixed values of $\rcut$.
In Figure~\ref{fig:r_cut} we investigate this behavior and show also the ($\rcut$ independent) NLO result obtained with \Matrix, by using dipole subtraction, and \madgraph, by using FKS subtraction.
As expected, the $\rcut$ dependence is {\it linear}~\cite{Catani:2017tuc,Buonocore:2019puv}, contrary to what happens in the case of the production of a colourless final-state system (see Sec.~7 of Ref.~\cite{Grazzini:2017mhc}), where the power-like dependence of the total cross section on $\rcut$ is known \cite{Grazzini:2016ctr,Ebert:2018gsn,Cieri:2019tfv}
to be quadratic (modulo logarithmic enhancements).

%%====================================
%%====================================

\begin{figure}[p]
\begin{center}
\includegraphics[width=0.49\textwidth]{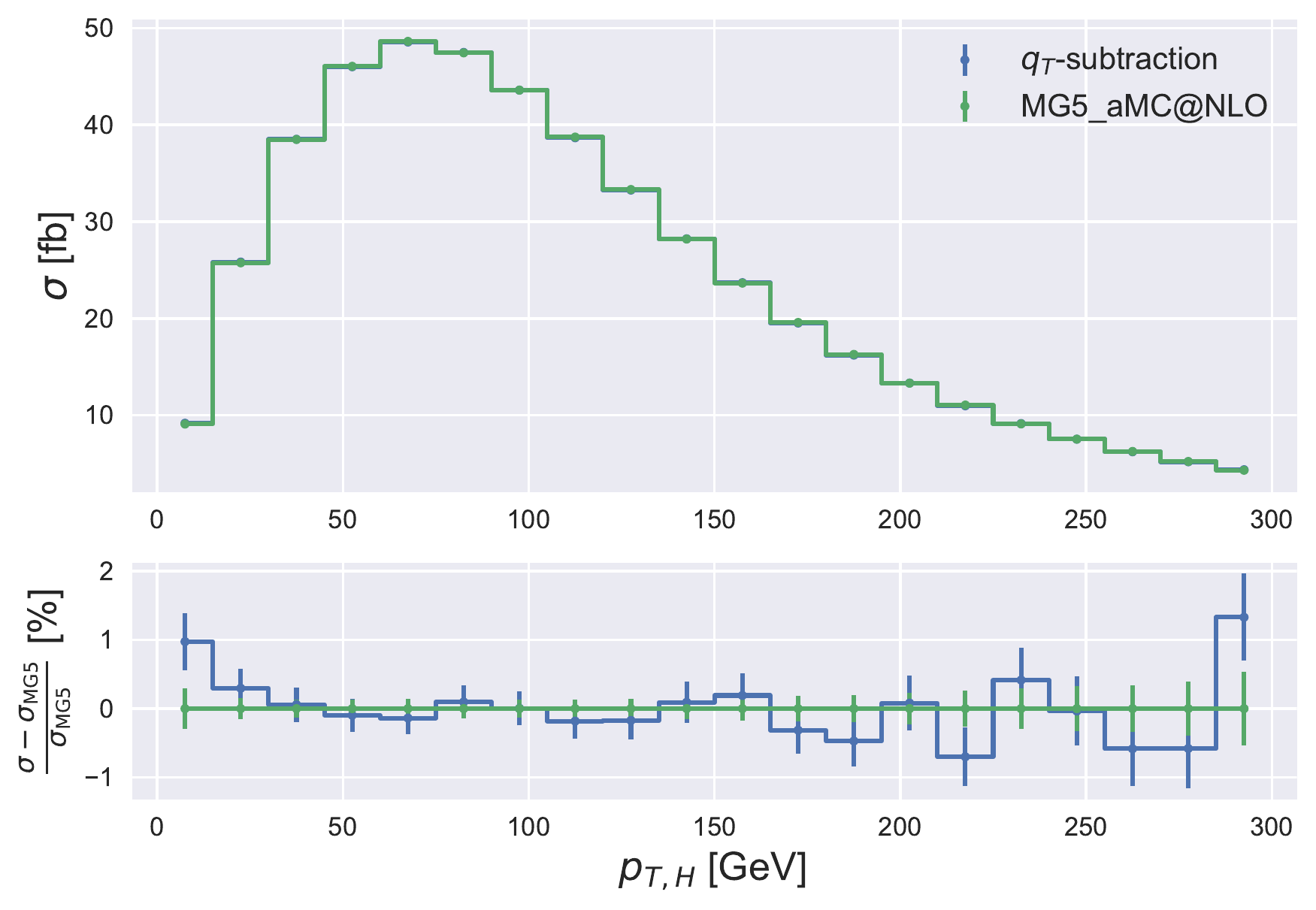}
\includegraphics[width=0.49\textwidth]{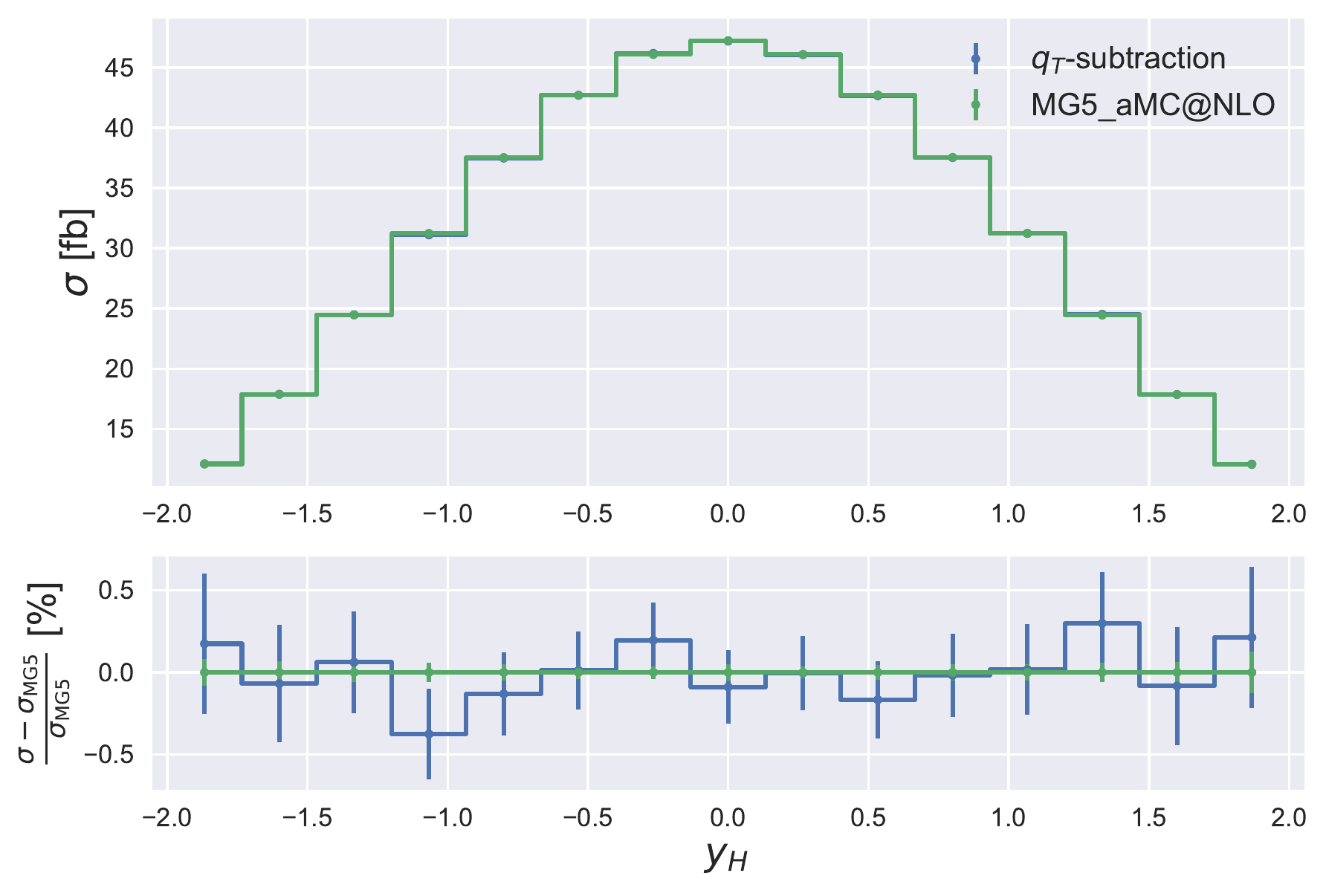}\\[2ex]
\includegraphics[width=0.49\textwidth]{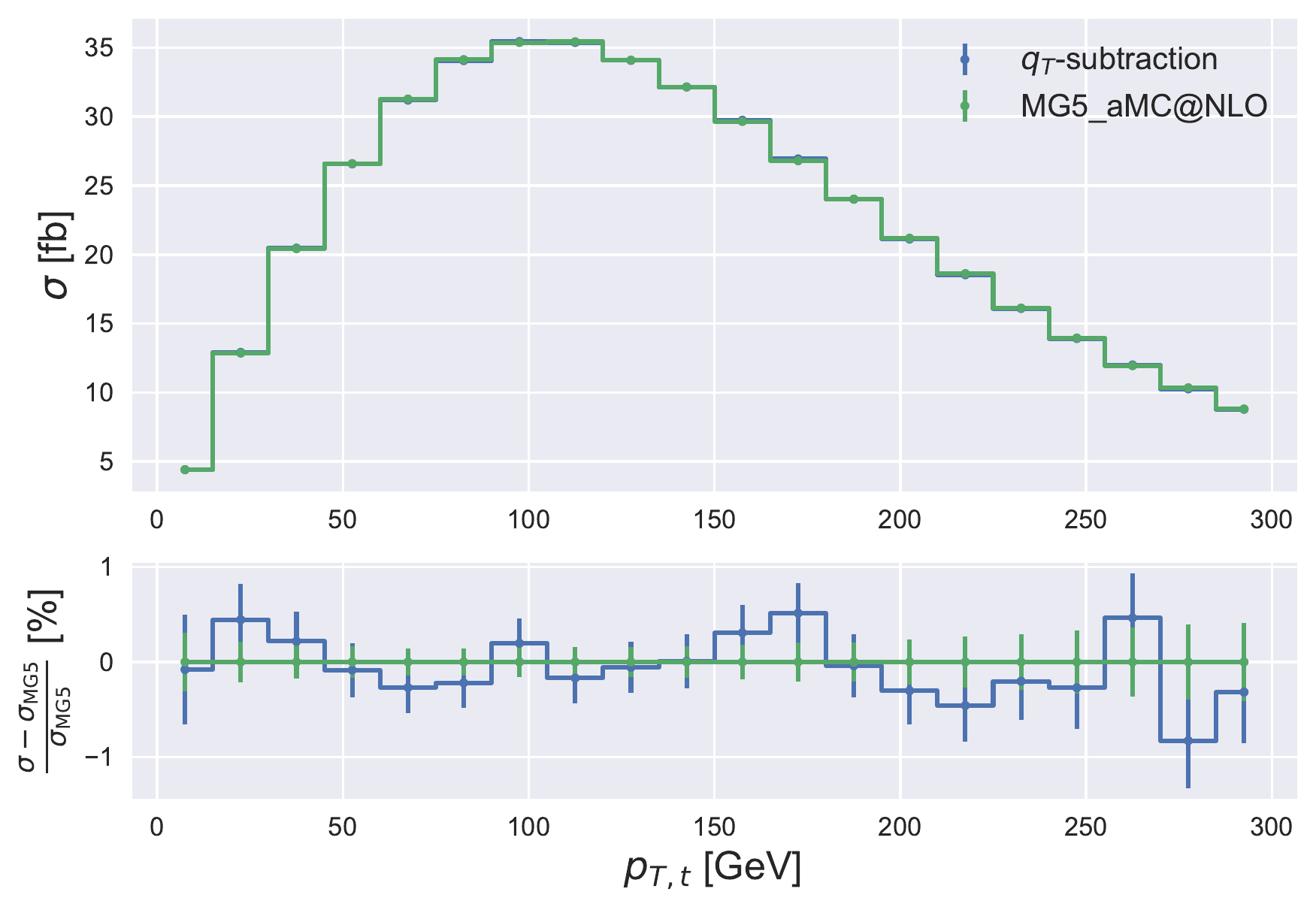}
\includegraphics[width=0.49\textwidth]{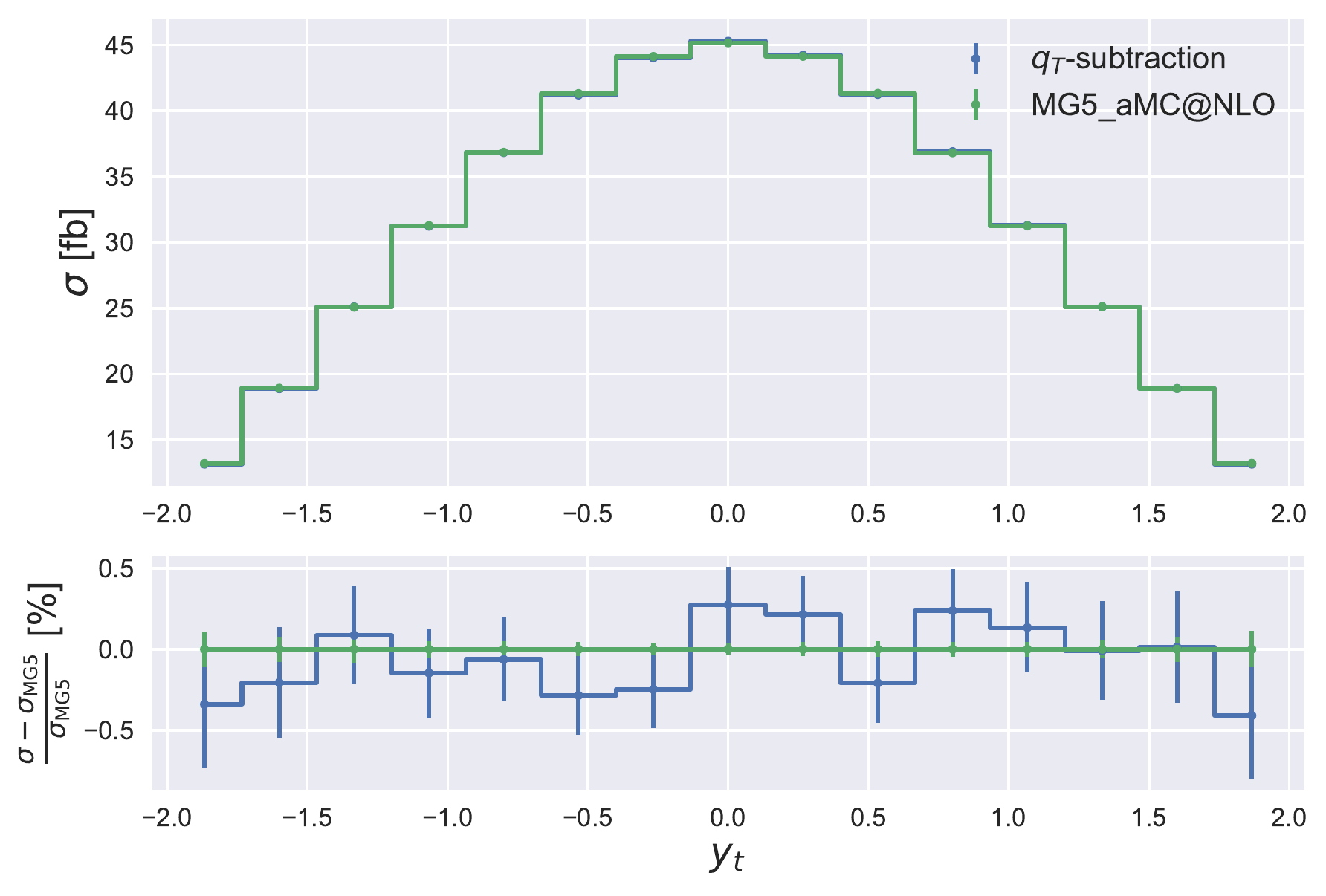}\\[2ex]
\includegraphics[width=0.49\textwidth]{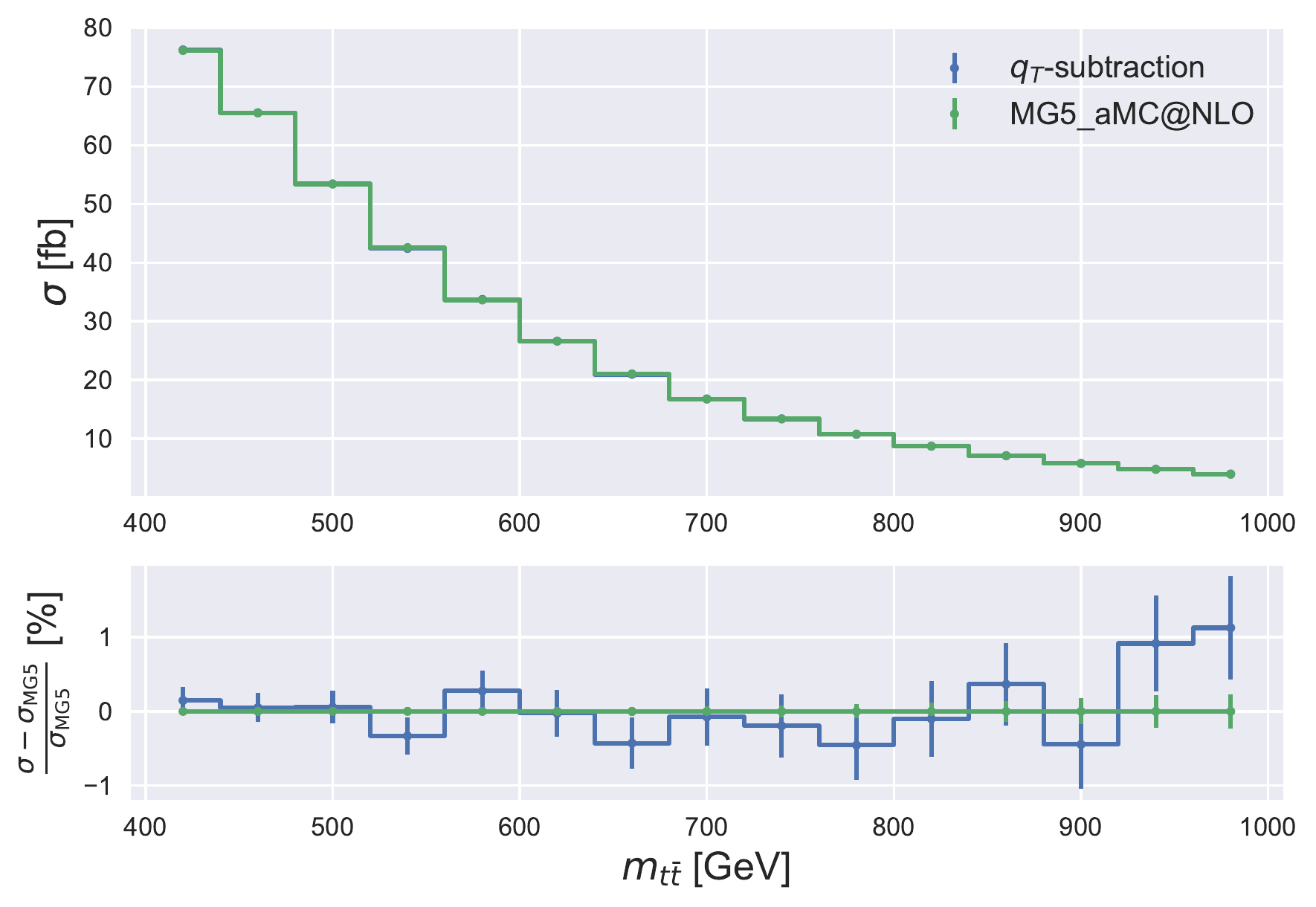}
\includegraphics[width=0.49\textwidth]{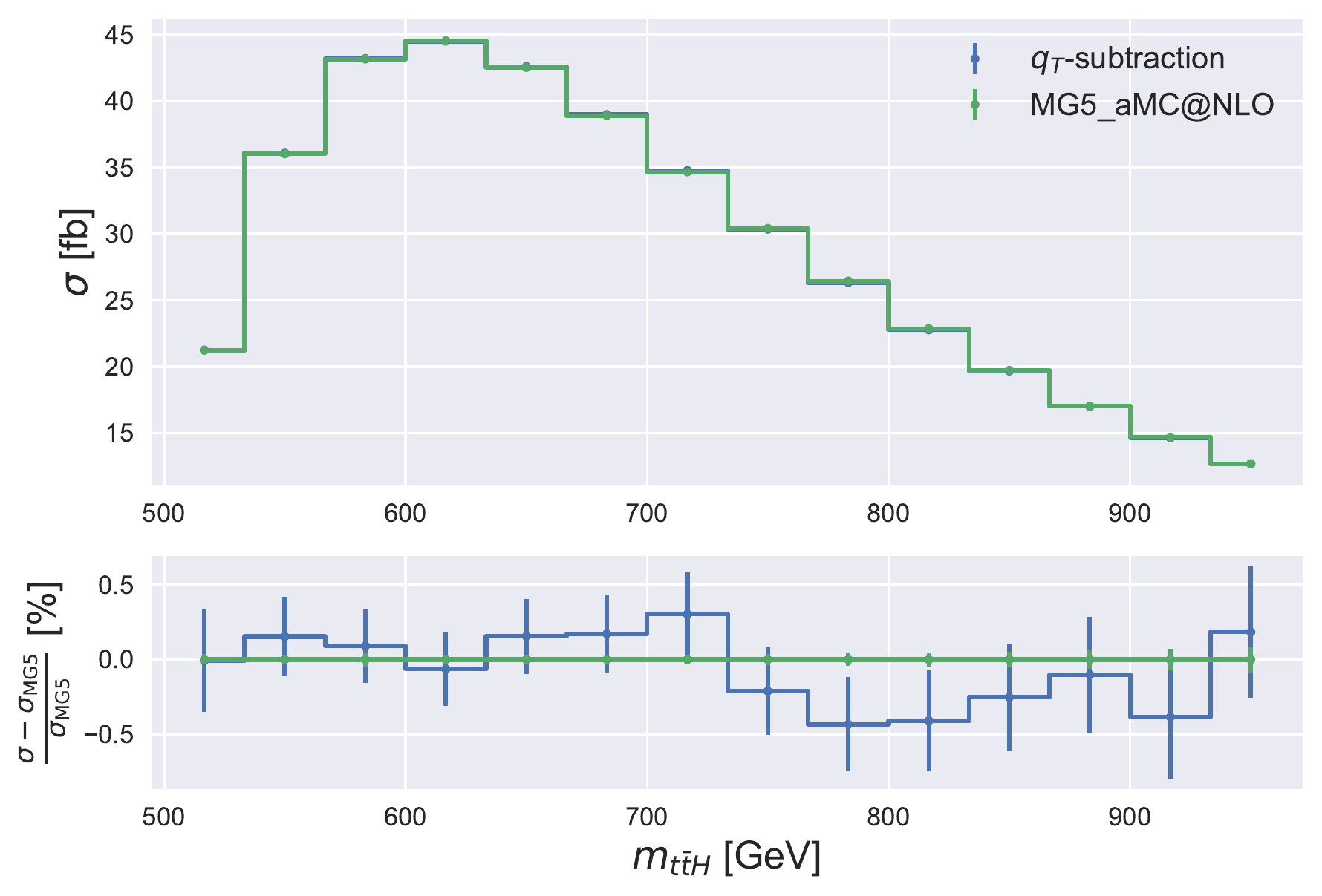}
\end{center}
\vspace{-1ex}
\caption{\label{fig:diff}
The NLO results of \madgraph for the cross section dependence on several kinematic variables at $\sqrt{s}=13~\mathrm{TeV}$. The lower panels show the relative comparison
with the corresponding results obtained by using $q_T$ subtraction.
}
\end{figure}

In Figure~\ref{fig:diff} we present the NLO results for several differential distributions at $\sqrt{s}=13~\mathrm{TeV}$ and compare them with those obtained by using \madgraph. 
In particular we consider the transverse-momentum (top left) and rapidity  (top right) distributions of the Higgs boson,
the transverse-momentum (center left) and rapidity  (center right) distributions of the top quark,
and the invariant-mass distributions of the top-quark pair (bottom left) and of the \ttbH system (bottom right).
We find excellent agreement between the two calculations, with bin-wise uncertainties at the percent-level or below.
Our results are obtained by using a fixed value of $\rcut$, $\rcut=0.1\%$, which, as suggested by Figure~\ref{fig:r_cut},
already provides a good estimate of the NLO cross section. A similar level of agreement is found with the results obtained with dipole subtraction.
We checked that the agreement also holds for different values of $\mu_R$ and $\mu_F$.

%%====================================
%%====================================   

We now move to considering the NNLO contributions to the total cross section.
In Table~\ref{tab:results} we report our results for the ${\cal O}(\as^4)$ contributions to the NNLO cross section from the flavour off-diagonal partonic channels $a_1a_2\to t{\bar t}H+X$.
The contribution from all the channels
with $a_1a_2= qg,{\bar q}g$ is labelled by the subscript $qg$, and the
contribution from all the channels with
$a_1a_2=qq, {\bar q}{\bar q}, qq', {\bar q}{\bar q}', q{\bar q}', {\bar q} q'$ $(q\neq q')$
is labelled by the subscript $q({\bar q})q^\prime$.
We see that the NNLO corrections from both contributions are very small, at the few {\it per mille} level of the NLO cross section. At $\sqrt{s}=13$ TeV they contribute with similar size and opposite sign,
and, therefore, their overall quantitative effect in this setup is completely negligible.
We also see that the numerical uncertainty of the NNLO correction in the $qg$ channel is rather large.
This is due to a cancellation between the two terms in Eq.~(\ref{eq:main}): the term ${\cal H}\otimes d{\hat \sigma}$, which is $\rcut$ independent, and the term in the square bracket, which depends on $\rcut$.
This cancellation is observed at both $\sqrt{s}=13$ TeV and $\sqrt{s}=100$ TeV, but it is particularly severe at $\sqrt{s}=13$ TeV, downgrading the numerical precision that can be obtained for the relative correction.
Similar effects were observed for $t{\bar t}$ production in Refs.~\cite{Bonciani:2015sha,Catani:2019iny}.

\begin{figure}[t]
\begin{center}
\includegraphics[width=0.48\textwidth]{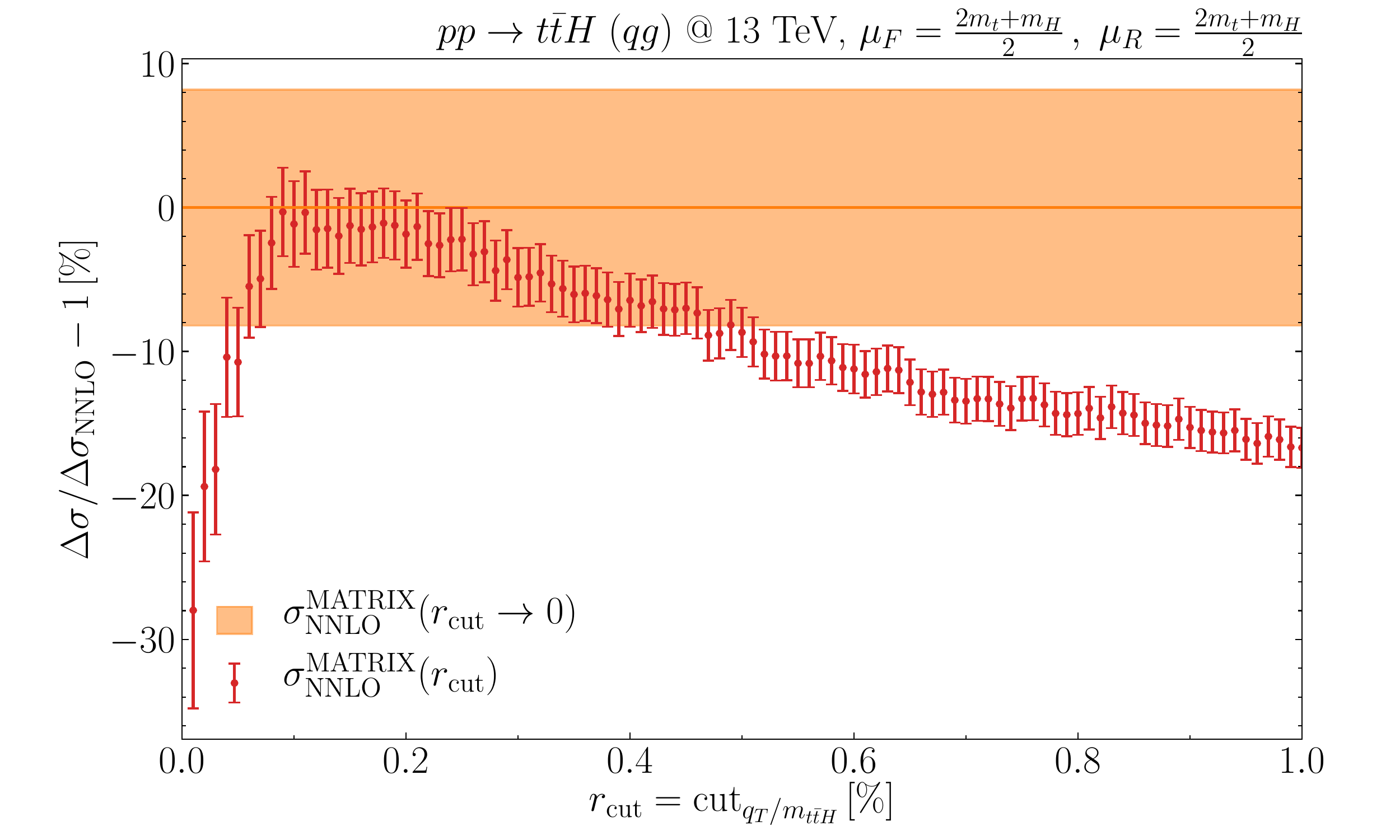}
\includegraphics[width=0.48\textwidth]{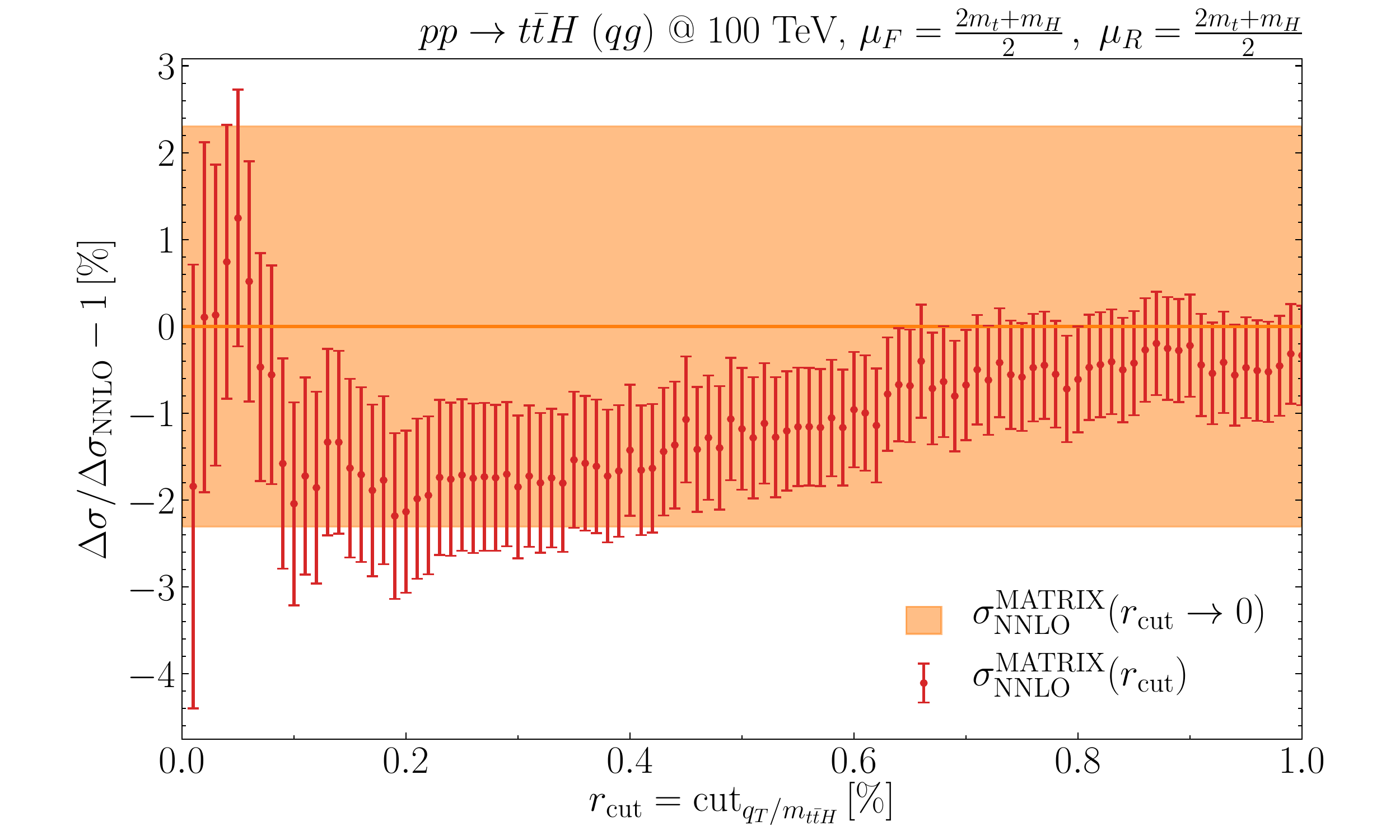}\\[2ex]
\includegraphics[width=0.48\textwidth]{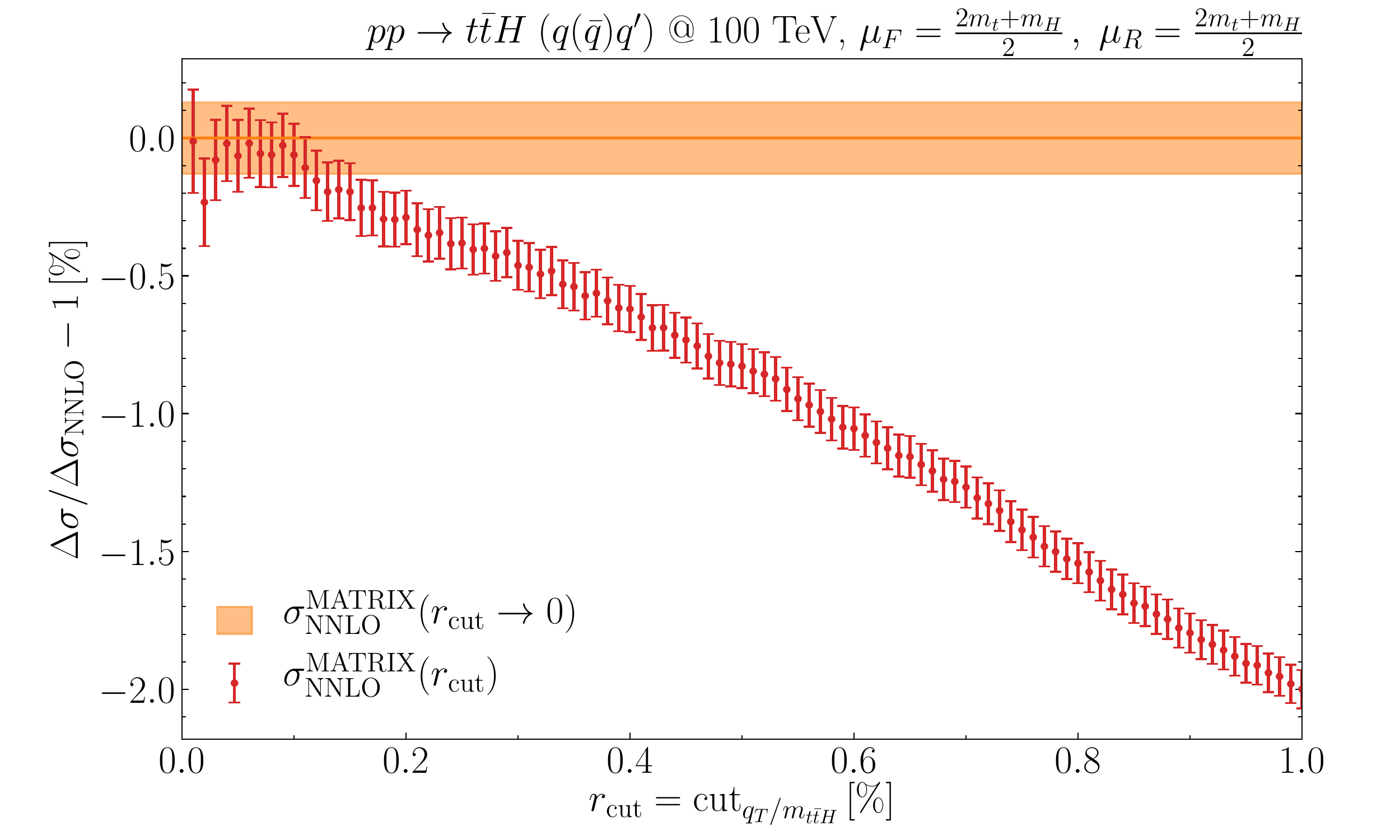}
\includegraphics[width=0.48\textwidth]{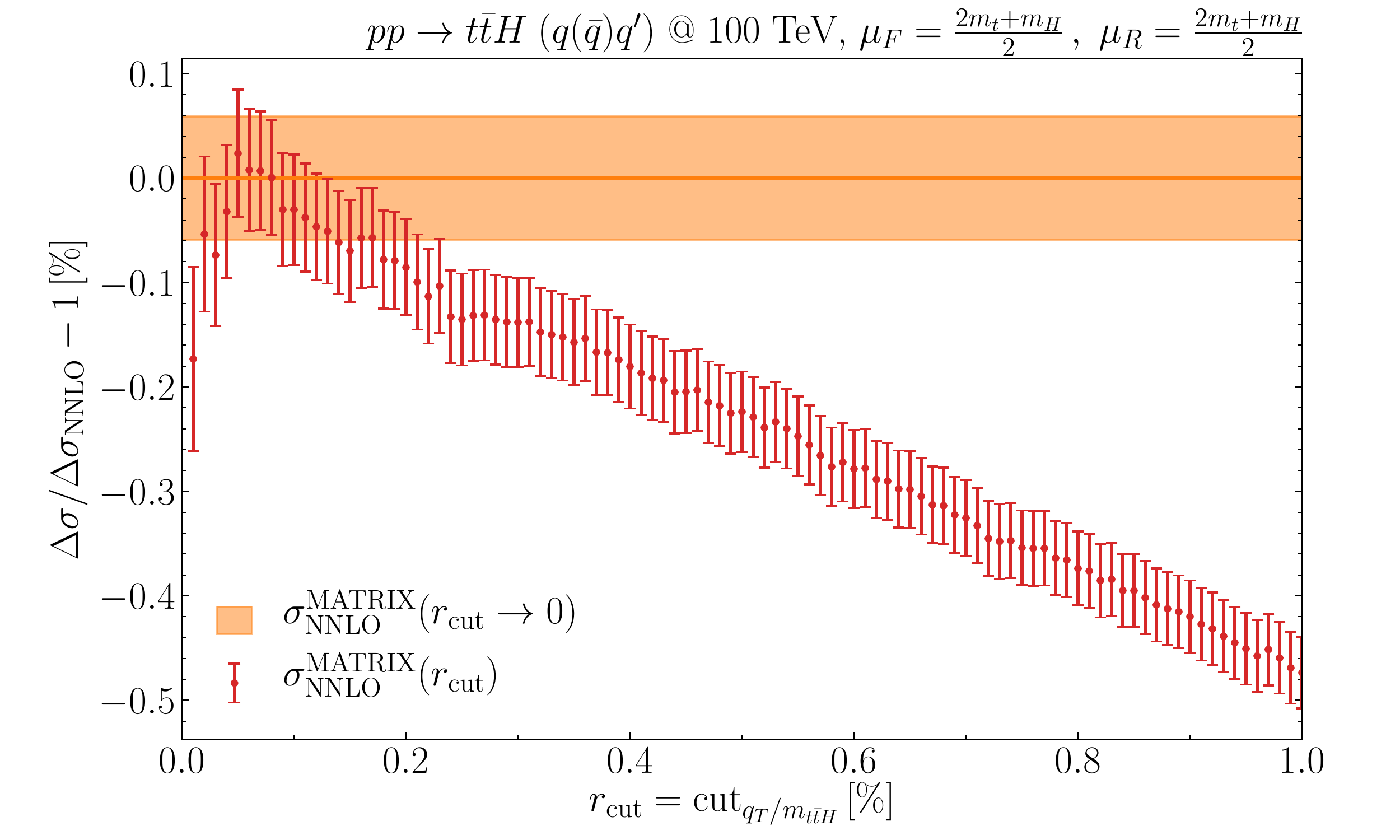}
\end{center}
\vspace{-1ex}
\caption{
The NNLO contribution $\Delta \sigma$ of the $qg$ (top) and $q({\bar q})q^\prime$ (bottom) partonic channels to the total cross section at $\sqrt{s}=13~\mathrm{TeV}$ (left) and $100~\mathrm{TeV}$ (right). The $\rcut$ dependence of $\Delta \sigma$ is normalized to its
extrapolation $\Delta \sigma_{\rm NNLO}$ at $\rcut \to 0$.}
\label{fig:nnlo}
\end{figure}

In Figure~\ref{fig:nnlo} we show the $\rcut$-dependence of the NNLO corrections $\Delta\sigma$ of the flavour off-diagonal channels to the total cross section for $\ttbH$ production.
The result is normalised to our extrapolation $\Delta\sigma_{\rm NNLO}$ at $\rcut\to 0$.
In the $qg$ channel at $\sqrt{s}=13$ TeV the extrapolation is particularly delicate, due to the cancellation discussed above.
For some of the channels, the first few points at low $\rcut$ values show relatively large instabilities, in particular
again for the $qg$ channel at $\sqrt{s}=13$ in Fig.~\ref{fig:nnlo} (top-left).
However, these points are not dropped in the fit, which is dominated by the behaviour at $\rcut>0.1\%$.
The $\rcut$ dependence confirms that our calculation can control the NNLO contributions in the off-diagonal partonic channels at the few percent level.
Comparing with the $r_{\rm cut}$ behaviour for $t{\bar t}$ production (see Fig. 1 of
Ref.~\cite{Catani:2019iny}), the behaviour in Fig.~\ref{fig:nnlo} for $t{\bar t}H$ production is qualitatively and quantitatively similar, and we do expect
the extrapolation at $r_{\rm cut} \to 0$ in the flavour diagonal channels to work
in a similar way for both production processes.
Based on the experience with the NNLO calculations for heavy-quark production \cite{Catani:2019iny,Catani:2019hip,Catani:2020kkl},
this should be fully sufficient to obtain
precise NNLO results by using the $q_T$ subtraction method once the presently unknown soft contributions and the two-loop amplitudes become available.

%%====================================
%%====================================

\section{Summary}
\label{sec:summa}

In this paper we have considered the associated production of the SM Higgs boson with a top-quark pair, and, more generally, processes in which heavy-quark pairs are produced in association with a colourless final-state system $F$.
We have pointed out that the transverse-momentum resummation formalism developed for $Q{\bar Q}$ production
in Ref.~\cite{Catani:2014qha} can be extended to associated $Q{\bar Q}F$ production.
This extension, which requires the evaluation of the appropriate resummation coefficients at the necessary perturbative accuracy, is also sufficient to apply the $q_T$ subtraction method to this class of processes.

Using the resummation coefficients presented in this paper and the current knowledge of scattering amplitudes,
it is possible to apply the $q_T$ subtraction formalism to $Q{\bar Q}F$ production up to NLO
and to obtain the NNLO corrections in all the flavour off-diagonal partonic channels.

We have implemented for the first time the $q_T$ subtraction formalism for \ttbH production, and we have presented first quantitative results at NLO and NNLO.
The calculation is accurate at NLO in QCD, and the NNLO corrections have been computed for the flavour off-diagonal partonic channels.
At NLO we have checked the correctness of our implementation by comparing with the results obtained by using tools that are based on established subtraction methods.
We found complete agreement for the total cross section and for single-differential distributions.
Within the setup that we have considered, we have found that the NNLO contribution of the off-diagonal partonic channels to the total cross section has a very small quantitative effect.
The extension of this calculation to the diagonal channels requires 
further theoretical work to compute the two-loop virtual amplitudes, and the NNLO soft contributions to the resummation coefficients.

\vskip 0.5cm
\noindent {\bf Acknowledgements}    

\noindent We are grateful to Federico Buccioni, Jean-Nicolas Lang, Jonas Lindert and Stefano Pozzorini
for their continuous assistance on issues related to \OpenLoops during the course of this project, and for providing us with the specific spin- and colour-correlated amplitudes which were necessary to complete the calculation.
This work is supported in part by the Swiss National Science Foundation (SNF) under contracts IZSAZ2$\_$173357 and 200020$\_$188464. The work of SK is supported by the ERC Starting Grant 714788 REINVENT.

\section*{Appendix}

In this Appendix we report the explicit expressions~\cite{unpub} of the resummation coefficients relevant for the production of an arbitrary number of heavy quarks $Q_j$ (with colour charges ${\bf T}_j$, pole masses $m_j$ and transverse momenta ${\bf p}_{jT}$) accompanied by a colourless system $F$ with total momentum $p_F$. More precisely, we consider the partonic process
\begin{equation}
\label{eq:qqfpro}
c(p_1) {\bar c}(p_2) \to Q_3(p_3) {\bar Q}_4(p_4) \dots Q_{N-1}(p_{N-1}) {\bar Q}_{N}(p_N) F(p_F)\, .
\end{equation}
where the final-state system is produced by either massless $q{\bar q}$ annihilation ($c=q$) or $gg$ fusion ($c=g$).
All the momenta $p_i$ ($i \geq 1$) are in the physical region, i.e. four-momentum conservation reads $p_1+p_2=p_3+...p_N+p_F$, while colour is outgoing, such that $\sum_{i=1}^N {\bf T}_i=0$. In particular we have ${\bf T}_1^2={\bf T}_2^2= C_c$, where $C_c=C_F$ if $c=q$ and
$C_c=C_A$ if $c=g$.
This notation corresponds to that used in Ref.~\cite{Catani:2014qha} for the simpler case of heavy-quark pair production,
$c(p_1) {\bar c}(p_2) \to Q(p_3) {\bar Q}(p_4)$.
We note that the massless and heavy quarks and antiquarks of the process in Eq.~(\ref{eq:qqfpro}) can have different flavours
(e.g., we can have $u {\bar d} \to t {\bar b} F$).

The coefficients that are presented below refer to perturbative expansions (see Eqs.~(\ref{eq:softan}), (\ref{eq:Dall}) and (\ref{eq:itilall})) in which $\as(\mu^2)$ denotes the renormalised QCD coupling
in the $\msbar$ scheme with decoupling of the heavy quarks $Q_j$ \cite{Bernreuther:1981sg}. In the case of $\ttbH$ production $\alpha(\mu^2)$ is the QCD coupling in the five-flavour scheme.

The first-order soft anomalous dimension $\g^{(1)}(\{p_i\})$ (see Eq.~(\ref{eq:softan})) reads
\begin{align}
\g^{(1)}(\{p_i\}) = -\f{1}{4}& 
\left\{ \sum_{j \geq 3} \T_j^2 \;(1 - i \pi)
+ \sum_{\substack{i=1,2 \\ j \geq 3}} 
\;\T_i \cdot \T_j \,\ln\frac{(2p_ip_j)^2}{M^2 m_j^2} \right. \nn\\
&+ \left. \sum_{\substack{j,k \geq 3 \\ j \neq k}}
\;\T_j \cdot \T_k \left[ \frac{1}{2\vjk} \ln\left(\frac{1+\vjk}{1-\vjk}\right) - 
i \pi \left( \frac{1}{\vjk} + 1 \right)
\right]
\right\},
\label{eq:gamma1}
\end{align}
where
\begin{align}
\vjk= \sqrt{1 - \frac{m_j^2 m_k^2}{(p_jp_k)^2}}\,.
\end{align}

The first-order subtraction operator $\widetilde{\bf I}^{(1)}$ (see Eq.~(\ref{eq:itilall})) can be written as 
\begin{align}
\label{eq:Itilde1}
\;\widetilde{\bf I}^{(1)}(\ep,M^2/\mu_R^2;\{p_i\})) 
&= - \frac{1}{2}
\left( \frac{M^2}{\mu_R^2}\right)^{-\ep} \left\{
\sum_{i=1,2} \left[ \left(\frac{1}{\ep^2} +i\pi \frac{1}{\ep} 
-\frac{\pi^2}{12}\right) \T_i^2 
+ \frac{1}{\ep} \,\gamma_i \right] \right. \nn \\
&\left. - \frac{4}{\ep} \;\g^{(1)}(\{p_i\}) + \F(\{p_i\}) \right\} \;,
\end{align} 
where the coefficients $\gamma_i$ ($i=q,{\bar q},g$)
originate from collinear radiation and
read $\gamma_q=\gamma_{\bar q}=3C_F/2$
and $\gamma_g= (11C_A-2N_f)/6$, $N_f$ being the number of flavours of
massless quarks.
The function $\F(\{p_i\})$ in Eq.~(\ref{eq:Itilde1}) 
is
\begin{align}
\label{eq:f1t}
\F(\{p_i\}) =
 \sum_{j \geq 3} \T_j^2 
\;\ln\left(\frac{m_j^2 + {\bf p}_{jT}^2}{m_j^2}\right)
- \sum_{\substack{i=1,2 \\ j \geq 3}} 
\;\T_i \cdot \T_j \;
{\rm Li}_2\left(-\frac{{\bf p}_{jT}^2}{m_j^2} \right)
+ \sum_{\substack{j,k \geq 3 \\ j \neq k}}
\;\T_j \cdot \T_k \;\frac{1}{2 \vjk} \,L_{jk} \;,
\end{align}
where
\begin{align}
\label{eq:ljk}
L_{jk}=&\; \frac{1}{2} \ln\left( \frac{1+\vjk}{1-\vjk} \right)
\, \ln \left[ 
\frac{(m_j^2+ {\bf p}_{jT}^2) (m_k^2+{\bf p}_{kT}^2 )}{m_j^2 \,m_k^2}
\right]
- 2 \,{\rm Li}_2\left( \frac{2 \vjk}{1+\vjk}\right)
- \frac{1}{4}\ln^2\left( \frac{1+\vjk}{1-\vjk} \right) \nn\\
&+ \sum_{i=1,2} \left[ \,{\rm Li}_2\left( 1 - 
\sqrt{\frac{1-\vjk}{1+\vjk}} \;r_{jk,i} \right)
+ \,{\rm Li}_2\left( 1 - 
\sqrt{\frac{1-\vjk}{1+\vjk}} \;\frac{1}{r_{jk,i}} \right)
+ \frac{1}{2} \ln^2 r_{jk,i} 
\right] 
\end{align}
with
\begin{equation}
r_{jk,i} \equiv \frac{m_k}{m_j} \frac{p_i \cdot p_j}{p_i \cdot p_k}\, .
\end{equation}
The function ${\rm Li}_2(z)$ in Eq.~(\ref{eq:ljk}) is the customary dilogarithm function,
${\rm Li}_2(z) = -\!\int_0^1 \frac{dt}{t} \ln(1-zt)$.

The coefficient $\dcor^{(1)}({\bf \hat b}, \{p_i\})$ (see Eq.~(\ref{eq:Dall})) reads
\begin{align}
\dcor^{(1)}({\bf \hat b}, \{p_i\}) =&\; \sum_{j \geq 3} \T_j^2 \left(
\frac{{\bf \hat b}\cdot {\bf p}_{jT}}{\sqrt {m_j^2 + 
\left( {\bf \hat b}\cdot {\bf p}_{jT}\right)^2}}\;
\left[ {\rm arcsinh}\left(\frac{{\bf \hat b}\cdot {\bf p}_{jT}}{m_j}\right)
+ \frac{i \pi}{2}\right]
- \frac{1}{2} \ln\left(\frac{m_j^2 + {\bf p}_{jT}^2}{m_j^2}\right)
\right)
\nn \\
&+ \sum_{\substack{i=1,2 \\ j \geq 3}} \T_i \cdot \T_j 
\left(
{\rm arcsinh}\left(\frac{{\bf \hat b}\cdot {\bf p}_{jT}}{m_j} \right)\!
\left[ {\rm arcsinh}\left(\frac{{\bf \hat b}\cdot {\bf p}_{jT}}{m_j}\right)
+ i \pi \, \right]\!
+ \frac{1}{2} {\rm Li}_2\!\left(-\frac{{\bf p}_{jT}^2}{m_j^2}\right)\!
\right)
\nn \\
&+\sum_{\substack{j,k \geq 3 \\ j \neq k}}
\T_j \cdot \T_k \;\frac{1}{2} \;D_{jk} \;\;,
\label{eq:d1}
\end{align}
with the function $D_{jk}$ that is given in terms of the following one-fold integral representation:
\begin{align}
D_{jk} = \int_0^1 dx \;\frac{p_j\cdot p_k}{w_{jk}^2(x)}
&\left(
\frac{2\,{\bf \hat b}\cdot {\bf w}_{jkT}(x)}{\sqrt {w_{jk}^2(x) + 
\left( {\bf \hat b}\cdot {\bf w}_{jkT}(x)\right)^2}}\;
\left[ {\rm arcsinh}\left(\frac{{\bf \hat b}\cdot {\bf w}_{jkT}(x)}{\sqrt
{w^2_{jk}(x)}}\right)
+ \frac{i \pi}{2}\right] \right. \nn\\
&\left.
- \ln\left(1+\frac{{\bf w}_{jkT}^2(x)}{w_{jk}^2(x)}\right)
\right) \;\;,
\label{eq:djkint}
\end{align}
where the four-vector $w_{jk}^\mu(x)$ is defined as
\begin{equation}
w_{jk}^\mu(x) = x p_j^\mu + (1-x) p_k^\mu\, .
\end{equation}

We note that the expressions of the resummation coefficients in 
Eqs.~(\ref{eq:gamma1}), (\ref{eq:Itilde1}), (\ref{eq:f1t}) and (\ref{eq:d1})
for the process in Eq.~(\ref{eq:qqfpro}) have no explicit dependence on the flavour of the
(massless and heavy) quarks and on the colourless system $F$. In particular, there is no dependence
on the quantum numbers of $F$ and on its momentum $p_F$ (though the dependence on $p_F$
enters implicitly through momentum conservation, $p_1+p_2=p_3+...p_N+p_F$).

Using Eqs.~(\ref{eq:gamma1}), (\ref{eq:Itilde1}), (\ref{eq:f1t}) and (\ref{eq:d1}),
we also note that we can recover the resummation coefficients of  
Ref.~\cite{Catani:2014qha} for the simpler process of heavy-quark pair 
production\footnote{In this case, the one-fold integral in Eq.~(\ref{eq:djkint}) is expressed 
through  dilogarithms in Eqs.~(36)--(38) of Ref.~\cite{Catani:2014qha}.},
$c(p_1) {\bar c}(p_2) \to Q(p_3) {\bar Q}(p_4)$. To this purpose we can simply use
the corresponding constraints $m_3=m_4, \T_1 + \T_2 = - ( \T_3 + \T_4)$ and, importantly,
the constraint $ {\bf p}_{3T} = - {\bf p}_{4T}$ that follows from momentum conservation
(i.e., $p_F=0$). Indeed, to a large extent, the main difference between heavy-quark pair production
and the process in Eq.~(\ref{eq:qqfpro}) is due to the fact that the $N$ transverse momenta
$ {\bf p}_{jT}$ ($3 \leq j \leq N$) in  Eq.~(\ref{eq:qqfpro}) are independent kinematical variables
(since ${\bf p}_{FT} \neq 0$). This main kinematical difference leads to technical computational complications
(e.g., in the analytic evaluation of the resummation coefficients) and to new dynamical features. 

One of these dynamical features regards azimuthal correlations. The ${\bf b}$-space resummation coefficient
$\dcor^{(1)}$ (which is due \cite{Catani:2014qha} to soft-parton radiation from the final state and 
from initial-state/final-state interferences) depends on the relative azimuthal angles
$\phi_{jb} = \phi({\bf p}_{jT}) - \phi({\bf b})$. This dependence produces ensuing azimuthal correlations
\cite{Catani:2014qha} with respect to the observable angles $\phi_{jq} = \phi({\bf p}_{jT}) - \phi({\bf q}_T)$
of the ${\bf q}_T$-differential cross section at $q_T \neq 0$. In the limit $q_T \to 0$ the azimuthal-correlation 
cross section is divergent order-by-order in QCD perturbation theory (transverse-momentum resummation
eventually makes the cross section finite \cite{Catani:2017tuc}).

In the case of heavy-quark pair production, $\dcor^{(1)}$ depends on $\cos^2 \phi_{3b}= \cos^2 \phi_{4b}$
\cite{Catani:2014qha}, and this implies that it produces azimuthal-correlation divergences only in the
case of {\it even} harmonics (i.e., harmonics with respect to
$\cos^k \phi_{3q} $ with $k=2,4,6,\dots$) \cite{Catani:2017tuc}.
This dependence of $\dcor^{(1)}$ is due to the kinematical relation $ {\bf p}_{3T} \simeq - {\bf p}_{4T}$
at $q_T \to 0$. 
In the case of the process in Eq.~(\ref{eq:qqfpro}), the transverse momenta $ {\bf p}_{jT}$ are
kinematically independent even if $q_T \to 0$, and it turns out that 
$\dcor^{(1)}$ depends on both $\cos^2 \phi_{jb}$ and $\cos \phi_{jb}$ (the dependence
on $\cos \phi_{jb}$ is due to the contributions that are proportional to `$i \pi$' in the right-hand side
of Eqs.~(\ref{eq:d1}) and (\ref{eq:djkint})). This dependence implies that the process in Eq.~(\ref{eq:qqfpro})
is affected by lowest-order (and higher-order) divergent azimuthal correlations for both
even and {\it odd} harmonics (i.e., harmonics with respect to
$\cos^k \phi_{3q} $ with $k=1,3,5,\dots$). A quite general discussion of azimuthal correlations at small $q_T$
is presented in Ref.~\cite{Catani:2017tuc}, where is also pointed out that lowest-order divergent odd harmonics
can occur in other processes, such as $V+{\rm jet}$ and dijet production.

We conclude this Appendix by recalling \cite{Catani:2014qha} that the second-order
term ${\bf \Gamma}_t^{(2)}$  of the soft anomalous dimension ${\bf \Gamma}_t$ 
(see Eq.~(\ref{eq:softan}))  for the process in Eq.~(\ref{eq:qqfpro})  is related to the corresponding term
of the anomalous dimension matrix ${\bf \Gamma}(\mu)$ that controls the QCD IR divergences
of scattering amplitudes with massive external particles 
\cite{Catani:2000ef, Mitov:2009sv, Ferroglia:2009ep, Ferroglia:2009ii, Mitov:2010xw}.
We have
\begin{equation}
\label{gtot}
\g(\as;\{p_i\}) = \frac{1}{2}
\;{\bf \Gamma}^{\,\rm sub.}(\as,\{p_i\}) -
\left(\frac{\as}{\pi}\right)^2 \;\frac{1}{4}
\left( \; \left[ \g^{(1)}(\{p_i\}) \,, \F(\{p_i\}) \right] +
\pi \beta_0 \F(\{p_i\})
\right) +
{\cal O}(\as^3)\, ,
\end{equation}
where $12 \pi \beta_0 = 11 C_A - 2 N_f$, $\g^{(1)}(\{p_i\})$ and $\F(\{p_i\}$ 
are given in Eqs.~(\ref{eq:gamma1}) and (\ref{eq:f1t}), 
and ${\bf \Gamma}^{\,\rm sub.}$ is given below.
The perturbative expansion
of the right-hand side of Eq.~(\ref{gtot}) includes both the first-order and
second-order terms $\g^{(1)}$ and $\g^{(2)}$ (obviously,
${\bf \Gamma}^{\,\rm sub.}= 2 (\as/\pi) \g^{(1)}                                                                                 
+ {\cal O}(\as^2)$), while terms at ${\cal O}(\as^3)$ and beyond are neglected.
The `subtracted' anomalous dimension
${\bf \Gamma}^{\,\rm sub.}(\as;\{p_i\})$
is 
\begin{equation}
\label{gasub}
{\bf \Gamma}^{\,\rm sub.}(\as,\{p_i\}) =
{\bf \Gamma}(\mu) -
\left[ \frac{1}{2} (\T_1^2 + \T_2^2) \,\gamma_{\rm cusp}(\as)
\left( \ln \frac{M^2}{\mu^2} - i \pi \right) + 2 \gamma^c(\as) \right] \;\;,
\end{equation}
where the terms on the right-hand side are written by exactly using the
notation of Eq.~(5) of Ref.~\cite{Ferroglia:2009ii}.
The term ${\bf \Gamma}(\mu)$ is the anomalous-dimension matrix that controls the
IR divergences of the scattering amplitude
${\cal M}_{c{\bar c}\to Q_3 {\bar Q}_4 \dots F}$ for the process in Eq.~(\ref{eq:qqfpro}), 
while the square-bracket term on the
right-hand side of Eq.~(\ref{gasub}) is the corresponding expression of
${\bf \Gamma}(\mu)$ for a generic process $c{\bar c}\to F$, where
the system $F$ is colourless. 
The expressions of ${\bf \Gamma}(\mu), \gamma_{\rm cusp}(\as)$ and
$\gamma^c(\as)$ up to ${\cal O}(\as^2)$ are explicitly given in Ref.~\cite{Ferroglia:2009ii}.

\bibliography{biblio}

\end{document}